\documentclass[aps,prl,twocolumn]{revtex4_mod}
\usepackage{graphicx,amsmath}

\begin{document}

\title{Chemo-dynamical simulations of dwarf galaxy evolution} 
  \author{Simone Recchi} \affiliation{Department of Astrophysics, Vienna
  University,\\ T\"urkenschanzstrasse 17, 1180, Vienna, Austria\\
  E-mail address: simone.recchi@univie.ac.at} \date{\today}

\begin{abstract}
  In this review I give a summary of the state-of-the-art for what
  concerns the chemo-dynamical numerical modelling of galaxies in
  general and of dwarf galaxies in particular.  In particular, I focus
  my attention on $(i)$ initial conditions; $(ii)$ the equations to
  solve; $(iii)$ the star formation process in galaxies; $(iv)$ the
  initial mass function; $(v)$ the chemical feedback; $(vi)$ the
  mechanical feedback; $(vii)$ the environmental effects.  Moreover,
  some key results concerning the development of galactic winds in
  galaxies and the fate of heavy elements, freshly synthesised after
  an episode of star formation, have been reported.  At the end of
  this review, I summarise the topics and physical processes, relevant
  for the evolution of galaxies, that in my opinion are not properly
  treated in modern computer simulations of galaxies and that deserve
  more attention in the future.
%(No more than 150 words)
\end{abstract}
\maketitle

\section*{1. Introduction}
Galaxies are extremely complex astrophysical objects.  In order to
study the evolution of galaxies, a deep understanding of many physical
processes, covering a broad range of spatial and temporal scales, is
required.  On the smallest scales, electromagnetic radiation and
particle-particle and particle-radiation interactions determine the
thermal and ionisation status of the interstellar medium (ISM).  On
the largest scales, galactic winds and environmental effects
(interactions with neighbouring galaxies and with the intracluster
medium) regulate the mass budget of the galaxy and strongly affect
its metallicity.  Many other key physical processes such as star
formation, feedback, gas circulation and stellar dynamics operate on
intermediate spatial and temporal scales.

This review paper gives a summary of ingredients, methods, results and
challenges encountered in the study of the chemical and dynamical
evolution of galaxies, with particular emphasis on the study of dwarf
galaxies (DGs).  The main focus of this review is the theoretical
study of the chemo-dynamical evolution of galaxies by means of
computer simulations.  For a broader and more comprehensive summary of
properties and physical processes in galaxies, the book ``Dwarf
galaxies: keys to galaxy formation and evolution'' (Springer) can be
consulted.

The last three decades have seen an enormous surge of activity in the
study of DGs, the most numerous galaxy species in the Universe.
Advanced ground-based and space-born observatories have allowed the
observation of these faint objects in the local volume with incredible
detail.  From a theoretical perspective, the interest in the study of
DGs is manifold.  Their shallow potential well allows an easier
venting out of freshly produced metals than in more massive galaxies.
Thus, DGs are perhaps significant polluters of the intracluster and
intergalactic medium (\cite{klein12}, but see \cite{gm97}).  According
to the hierarchical scenario for galaxy formation, dwarf galaxy-sized
objects are the building blocks to form larger galaxies.  DGs do not
possess very prominent spiral structures or significant shear motions,
hence the study of the star formation in these objects is somewhat
easier than in spiral galaxies.

Besides providing key information about the kinematics of gas in
galaxies, spectroscopy allows the determination of the metallicity and
of the abundance ratios of specific elements.  This is a very useful
information because chemical abundances provide crucial clues to the
evolution of galaxies.  The increasing availability of large
telescopes made possible the systematic study of extragalactic
H\,{\small II} regions and other objects in external galaxies.  In
this way, variations of chemical composition between different
galaxies and in different positions within a single galaxy could be
studied.  Integral field spectroscopy in this sense is a fundamental
step forward.  Detailed maps of the chemical abundances within a
single galaxy can be obtained.  In order to understand the origin of
such distributions of metals, one often has to resort to the work and
models of theoreticians.

Although a few basic properties of galaxies can be understood with
simple analytical and semi-analytical considerations, the enormous
complexity of galactic physics can only be handled (in part) with the
help of numerical simulations.  This is especially true for what
concerns the chemical evolution of galaxies.  Simple closed-box models
\cite{tins80} can provide a first-order explanation for the global
metallicity in a galaxy, but the spatial distribution of metals can
not be addressed with these simplified tools.  On the other hand, due
to the large number of processes one has to take into account,
numerical simulations make generally use of results taken from other
research fields and combine them in such a way that a detailed
description of the evolution of galaxies can emerge.  The process of
simulating galaxies is thus analogous to the process of cooking.  To
prepare a culinary dish, ingredients must be accurately chosen, the
necessary equipment must be in place, a number of steps and operations
must be performed to combine the ingredients and some times a personal
touch is added and standard cookbook recipes are modified in order to
obtain a special effect.

For chemo-dynamical simulators of galaxy evolution, the main
ingredients are:
\begin{itemize}
\item the initial conditions
\item the set of equations to solve
\item a description of the star formation process
\item the mass distribution of newly born stars (the initial mass
  function or IMF)
\item a description of the chemical feedback from stars to gas
\item a description of the energy interchange processes between stars
  and gas.  There are many processes one might take into account but
  all of them are usually referred to as feedback.  This includes also
  feedback processes related to the presence of supermassive black
  holes and active galactic nuclei (AGN).  These kind of processes are
  usually dubbed AGN feedback.
\item a description of the interactions between the galaxy and the
  surrounding environment (galaxy-galaxy interactions, ram-pressure
  stripping due to an external inter-galactic medium, gas infall and
  so on)
\end{itemize}
In this review, I will consider in some detail some of these
ingredients and I will describe how they have been parametrised and
implemented in numerical simulations of galaxies.  Ingredients related
to the chemical evolution of galaxies will be treated with particular
care.  In the description of these ingredients, some personal bias
will be applied and higher priority will be given to the most relevant
ingredients for the simulation of DGs.  In particular, AGN feedback
will be only very briefly mentioned.

In the process of preparing a dish, the necessary equipment (pans,
pots and stove) must be in place and the quality of the equipment
affects the final outcome.  This is also true for the numerical
simulation of galaxies, where the main equipment is a computer.  More
often, a cluster of computers equipped with fast processors is
necessary.  Besides having a fast computer, appropriate algorithms and
sophisticated numerical methods must be in place in order to
efficiently solve the complex equations describing the evolution of
galaxies.  Some of these methods will be summarised in this review,
too.  Again, besides a very brief survey of most widely adopted
methods, specific tools required for the study of the chemo-dynamical
evolution of galaxies will be described with more care.

Numerical simulations always address specific issues in the evolution
of galaxies, trying to give answers to open problems or trying to
provide explanations to observed properties and characteristics of
galaxies or groups of galaxies.  In this review I will give a summary
of the state-of-the-art for what concerns some of these specific
issues.  In particular, I will focus on the conditions for the
development of galactic winds and on the fate of heavy elements,
freshly produced during an episode of star formation.

The organisation of this paper is thus quite simple: there is a
Section for each ingredient: initial conditions (Sect. 2), the
equations (Sect. 3), the star formation (Sect. 4), the initial mass
function (Sect. 5), the chemical feedback (Sect. 6), the mechanical
feedback (Sect. 7) and the environmental effects (Sect. 8).  In each
section, commonly adopted methodologies and recipes will be introduced
and some key results of past or ongoing studies will be summarised.
In Sect. 9 I will summarise some relevant results of numerical
investigations of DGs concerning galactic winds and their
consequences.  Finally, in Sect. 10 some conclusions will be drawn.

\section*{2. The initial conditions}
Nowadays it is pretty common to find in the literature studies of the
formation and evolution of galaxies in a cosmological context, meaning
that initial conditions consist of a scale-free or nearly scale-free
spectrum of Gaussian fluctuations as predicted by cosmic inflation and
with cosmological parameters determined from observations of the
cosmic microwave background radiation obtained by spacecrafts such as
WMAP \cite{sper03, koma11}.  However, the most detailed and
sophisticated cosmological simulations to date, such as the
Millennium-II simulation \cite{boyl09} and the Bolshoi simulation
\cite{klyp11} have force resolutions of the order of 1 kpc.  This is
barely enough to resolve large galaxies, but it is clearly
insufficient to resolve in detail DGs, whose optical radii are some
times smaller than that.  A lot in resolution can be gained by zooming
in and re-simulating small chunks of a large cosmological box
\cite{spri08, diem08, stin13}.  This method is gaining pace and has
been applied by various groups to DGs \cite{mart09, sawa11, pilk12}.
Still, at the present time the best way to accurately simulate a DG is
by numerically studying it as a single isolated entity \cite{maye06,
  stin07, revaz09, sb10, schr11, teys13}.

Numerical studies of galaxies in isolation assume some initial
configuration of gas density, temperature and stellar distribution.
This initial configuration is an equilibrium status of the system.
Starting from an equilibrium condition is clearly necessary in order
to pin down the effect of perturbing phenomena (star formation,
environmental effects, AGN feedback, and so on).  

A common strategy is to consider a rotating, isothermal gas in
equilibrium with the potential generated by a fixed distribution of
stars and/or of dark matter \cite{ti88, stt98, mf99}.  Rotating gas
configurations are usually better described by means of a cylindrical
coordinate system $(R, \phi, z)$.  Often, axial symmetry is assumed.
The relevant equation to solve in order to find the density
distribution of gas $\rho(R,z)$ is thus the steady-state (time
independent) Euler equation
\begin{equation}
\frac{1}{\rho}\nabla P + ({\mathbf v}\cdot \nabla){\mathbf v}=-\nabla \Phi,
\label{eq:euler}
\end{equation}
\noindent
where $P$ is the pressure, ${\mathbf v}$ is the bulk velocity of the
gas and $\Phi$ is the total gravitational potential.  In this
equation, only the component $v_\phi$ of the velocity must be
considered because it gives centrifugal support against the gravity.
Eq. \ref{eq:euler} in fact implies that the gravitational pull is
counter-balanced by the combined effect of pressure gradient and
centrifugal force.  

Most of the authors assume $\Phi$ to be independent of $\rho$.  This
means that the self-gravity of the gas is not considered.  A typical
justification of this choice is ``The omission of self-gravity is
reasonable, given that the baryonic-to-dark matter ratio of the
systems is $\sim$ 0.1.''  \cite{frag04}.  However, even if the total
mass of a DG is dominated by a dark matter halo, within the Holmberg
radius (the radius at which the surface brightness is 26.5 mag
arcsec$^{-2}$), most of the galaxy is made of baryons \cite{papa96,
  swat11}, so the inclusion of gas self-gravity in the central part of
a DG appears to be important.  I will come back to this point later in
this section.  For the moment it is enough to take note of the fact
that the assumption that $\Phi$ is independent of $\rho$ greatly
simplifies the calculation of the steady-state density configuration.
Furtermore, a barotropic equation of state $P=P(\rho)$ and a
dependence of the azimuthal velocity $v_\phi$ with known quantities is
commonly assumed.  

A widely used strategy is to assume that $v_\phi=ev_{circ}$, where
$v_{circ}=\sqrt{R\frac{d\Phi}{dR}}$ is the circular velocity and $e$
is the spin parameter that determines how much the galaxy is supported
against gravity by rotation and how much it is supported by the
pressure gradient.  A typical value for $e$ is 0.9, independent on the
height $z$ \cite{tb93, such94}.  \cite{ss00} assume that $e=0.9$ in
the plane of the galaxy, but it drops exponentially with height in
order to have non-rotating gas halos.  It is however important to
remark that, according to the Poincare'-Wavre theorem \cite{lebo67,
  tass80, barn06}, the rotation velocity of any barotropic gas
configuration (including thus also isothermal configurations) in
rotating equilibrium must be independent of $z$.  In other words, it
is possible to construct a centrifugal potential to add to $\Phi$ in
Eq. \ref{eq:euler} only if the circular velocity is independent on
$z$.

Other authors \cite{db99} solve instead the equilibrium equation in
the plane:
\begin{equation}
  v_\phi^2=v_{circ}^2-\frac{R}{\rho}\left| \frac{dP}{dR}
\right|_{z=0},
\end{equation}
\noindent
and assume the azimuthal velocity to be independent of $z$, in
compliance with the Poincare'-Wavre theorem.  The density at any $z$
is then found integrating the $z$-component of the hydrostatic
equilibrium equation, for any $R$.  Some authors then \cite{kauf07,
  dt08, teys13} set the gas in rotation around the $z$-axis, using the
average angular momentum profile computed from cosmological
simulations \cite{bull01}.

A different approach is followed by \cite{ab04}. Initially, there is
no balance between gravity and pressure and the gas collapses into the
midplane.  Supernovae (SNe) go off, principally along the disk and
this drives the collapsed gas upwards again.  Eventually, upward and
downward flowing gas come into dynamical equilibrium.  Some
multi-phase simulations \cite{hth06, liu13} adopt a similar approach
for the diffuse component, i.e. the distribution of diffuse gas starts
far from equilibrium. Then, it relaxes on a few dynamical time scales
to a quasi-equilibrium state, which represents the initial conditions
for the simulation.

One should be aware of the limitations of an equilibrium model without
gas self-gravity.  Most of the numerical simulations treat
self-consistently the process of star formation.  Since star formation
occurs when the gas self-gravity prevails over pressure, neglecting
the gas self-gravity in the set up of the model is clearly
inconsistent.  Moreover, without self-gravity, there is the risk of
building gas configurations which would have never been realized if
self-gravity were taken into account.  In order to solve these
problems, Vorobyov et al.  \cite{vrh12} explicitly took into account
gas self-gravity to build initial equilibrium configurations.  The
gravitational potential $\Phi$ is composed of two parts, one is due to
a fixed component (dark matter and eventually also old stars), one
($\Phi_g$) is due to the gas self-gravity. The gas gravitational
potential $\Phi_g$ is obtained by means of the Poisson equation
\begin{equation}
\nabla^2\Phi_g=4 \pi G \rho.
\label{Possion}
\end{equation}
\noindent
The gas density distribution is thus used to calculate the potential,
but this potential is then included in the Euler equation to find the
gas distribution.  Clearly, an iterative procedure, analogous to the
classical self-consistent field method \cite{om68}, is necessary to
converge to an equilibrium solution.

For a given mass $M_{\rm DM}$ of the dark matter halo, many solutions
are possible, according to the initial assumption about the density
distribution of the gas.  However, the self-gravitating equilibrium
configurations always have a maximum allowed gas mass $M_{\rm max}$,
unlike the case of non-self-gravitating equilibria which can realize
configurations with unphysically high gas masses.  Moreover, only for
some of the solutions, star formation was found to be permissible by
Vorobyov et al. (two star formation criteria based on the surface gas
density and on the Toomre parameter were assumed). The minimum gas
mass $M_{\rm g}^{\rm min}$ required to satisfy the star formation
criteria was found to be mainly dependent on the gas temperature
$T_{\rm g}$, gas spin parameter $e$ and degree of non-thermal support.
$M_{\rm g}^{\rm min}$ was then compared with $M_{\rm b}$, the amount
of baryonic matter (for a given $M_{\rm DM}$) predicted by the
$\Lambda$CDM theory of structure formation.  Galaxies with $M_{\rm
  DM}\geq 10^{9}~M_\odot$ are characterised by $M_{\rm g}^{\rm
  min}\leq M_{\rm b}$, implying that star formation in such objects is
surely possible as the required gas mass is consistent with what is
available according to the $\Lambda$CDM theory.  On the other hand,
models with $M_{\rm DM}\leq 10^{9}~M_\odot$ are often characterised by
$M_{\rm g}^{\rm min}\gg M_{\rm b}$, implying that they need much more
gas than available to achieve a state in which star formation is
allowed.  In the framework of the $\Lambda$CDM theory, this implies
the existence of a critical dark matter halo mass below which the
likelihood of star formation drops significantly (\cite{vrh12}).  It
is observationally well established that the galactic stellar mass
function for low-mass galaxies is quite shallow (i.e. $dn/dM_* \propto
M_*^\alpha$, with $\alpha \sim -1.3$, see \cite{cole01, ymb09}).  This
is at variance with the steeper ($\alpha \sim -1.8$) halo mass
function predicted over the mass range of interest by the $\Lambda$CDM
theory.  It seems thus that the efficiency of forming stars within
each dark matter halo decreases with the mass of the halo.  The
results of Vorobyov et al. illustrated above agree with this result
(see also \cite{ymb09, papa12, sm12}).

\section*{3. The equations}
In order to follow the evolution of a galaxy, the basic equations to
solve are of course the Euler equations, namely the standard set of
equations (conservation of mass, momentum and energy) governing
inviscid flows.  Viscosity in astrophysical plasmas is in fact usually
very small.  It can be large in some localised system, for instance in
accretion disks, but on a larger, galactic-wide scale the ISM can be
considered inviscid and there is no need to invoke the Navier-Stokes
equations.  Conversely, astrophysical plasmas are usually very
turbulent \cite{es04}.  In spite of that, also the use of turbulence
models in simulations of galaxies is still quite limited.  The main
reason for that is the lack of a satisfying characterisation and
modelling strategy for the compressible turbulence.  Progress in this
field is however constant and very sophisticated turbulence models
have been applied recently to astrophysical problems \cite{bd02,
  bran03, sb08, iapi08, glov10, micic12, iapi13}.  Important first
steps have been performed also in the simulation of turbulent gas in
galaxies \cite{ko09, sb10, bour10, mg12, rena13}.

Since a large volume fraction of the ISM of star forming galaxies is
ionised, a description of the electro-magnetic interactions is clearly
required.  This is most often realized by means of the so called ideal
magneto-hydrodynamical equations, where various ions are treated as a
single fluid, the conductivity of the ionised gas is assumed to be
very large and the plasma is assumed to be frozen in the magnetic
field.  Many modern hydrodynamical codes, such as ZEUS \cite{sn92},
FLASH \cite{fryx00, linde02}, RAMSES \cite{teys02, fht05}, ATHENA
\cite{stone08}, just to name a few, solve the ideal
magneto-hydrodynamical equations.  The inclusion of magnetic fields
affects the dynamics of gas in a galaxy in many ways. $(i)$ Magnetic
fields strongly reduce the transverse flow of charged particles, hence
the thermal conduction in directions orthogonal to field lines
\cite{spit62}.  Thermal conduction along field lines remains unaltered
compared to non-magnetised gases.  $(ii)$ Magnetic tension forces tend
also to suppress dynamical instabilities parallel, but not
perpendicular, to field lines \cite{ds09}.  Magnetic fields might also
inhibit the break-out of hot bubbles and superbubbles \cite{kama98}.
Also the mixing between the hot bubble and the surrounding cold
supershell can be reduced due to the presence of magnetic fields.
$(iii)$ The magnetic pressure $B^2/8\pi$ plays an important role in
the gas dynamics.  It is in fact comparable with the thermal pressure
and, if the magnetic field is not too weak, it is the dominant form of
pressure for temperatures below $\sim$ 200 K \cite{ab05}.  This is
consistent with the fact that the estimated thermal pressure in the
Milky Way is $\sim$ $5 \cdot$10$^{-13}$ dyne cm$^{-3}$, whereas the
estimated magnetic pressure is $\sim$ 10$^{-12}$ dyne cm$^{-3}$ (see
\cite{tiel05}).  Simulations of the formation of spiral galaxies
\cite{ps13} show indeed that the additional pressure due to magnetic
fields can lead to lower star formation rates at late times compared
to simulations without magnetic fields.  Also the structure of the
spiral arms is affected by the presence of magnetic fields.

It is less easy to assess the importance of magnetic fields in the
simulations of DGs.  In fact, not so much is known about magnetic
fields in these objects.  Starbursting DGs such as NGC1569
\cite{kepl10} or NGC4449 \cite{chyz00} are known to have magnetic
fields with strengths as high as few tens of $\mu$G, whereas quiescent
DGs have much weaker magnetic fields (a few $\mu$G, \cite{klein08,
  klein12}).  Magnetic fields are probably not the main drivers of DG
evolution, at least during periods of quiescent or weak star
formation.

Since our knowledge of galaxies almost exclusively depends on their
emitted (or absorbed) radiation, radiation hydrodynamics clearly
allows a description of galaxies which is more complete and easier to
compare with observations.  The radiation hydrodynamical equations are
more complex than the Euler equations.  A few textbooks exist, in
which these equations and related numerical methods are described in
detail \cite{mm84, kalk88, cast04}.  Many authors who attempted to
solve them made simplifying assumptions about the matter-radiation
coupling.  

The simplest possible way to include the effects of radiation in
hydrodynamical simulations is to assume that the gas is optically
thin.  The only effect of radiation is thus to reduce the available
thermal energy of the gas, i.e. radiation acts only as an energy sink.
Many works in the literature are devoted to the calculation of the
cooling function of an optically thin plasma \cite{bh89, sd93, schu09}
and these functions are used to calculate the rate of thermal energy
loss as a function of density, temperature and chemical composition.
A further commonly adopted assumption is the on the spot approximation
\cite{spit78}, according to which the photons produced in
recombination processes do not propagate but are immediately absorbed
locally.  In this way, the transport of these photons must not be
considered and the equations to solve simplify considerably.  The heat
produced by the radiation is transported out according to a law
similar to the thermal conduction.  This approximation turns out to be
valid as long as the particle density is sufficiently high, i.e.  when
the optically thick limit applies.  There are various examples of
radiation hydrodynamical simulations which make use of the on the spot
approximation \cite{lucy77, fhy03, fhy06, vbc06, grit09}.  A step
forward is the so called flux limited diffusion, where the optically
thin and optically thick limits are connected by appropriate flux
limiter functions \cite{wb06, frw06, kt12}.  Radiation hydrodynamics
is clearly very relevant and might quite substantially change our
understanding of galaxy formation and evolution of galaxies
\cite{co93, wise12, kim13b}.  In particular, the inclusion of
radiation feedback (photo-heating and radiation pressure) turns out to
be very important and it helps reproducing the observed distribution
of stellar masses in DGs, whereas simulations with only supernova
feedback fail to reproduce the observed stellar masses (\cite{hopk13},
see also Sect. 7).  In spite of significant recent progresses, the
inherent complexity has so far limited the use of radiation
hydrodynamical equations in galaxy simulations.

Of course, gas is not the only component of a galaxy.  Stars and, very
often, dark matter must be considered, too.  The gravitational
potential they generate has been already considered in Sect. 2.
However, their dynamics can be very important, as well.  The relevance
of a live dark matter halo for the evolution of a galaxy is not clear
and many authors still assume a fixed dark matter halo.  Conversely,
it is clear that the stellar dynamics plays an important role in the
evolution of a galaxy, at least if one is interested in time spans
larger than a few tens of Myr.  This has been demonstrated for
instance by Slyz \cite{slyz07} by means of a clear numerical
experiment.  According to this study, spurious results can be obtained
if one does not allow stars to move from their natal sites.  In
particular, the energy of Type II Supernovae (SNeII) is, in this case,
always released in regions of high densities (because in these regions
it is more likely to form stars, see Sect. 4), where cooling rates are
high.  This leads to the so-called overcooling problem (see also Sect.
7).  This problem can be simply avoided if one allows stars to move
during their lifetimes and, hence, SNeII to explode in environments
other than their natal ones (in particular, to explode in less dense
environments).

A widely used strategy to follow the dynamics of stars (and of dark
matter particles) is to consider individual stars, or more often,
populations of stars, as point masses and to follow their orbits by
means of standard N-body integration techniques.  This approach is
straightforward in SPH simulations of galaxies but it is widely used
also in grid-based codes.  However, in grid-based codes there is the
problem that star particles must be mapped to the mesh in order for
the global gravitational potential to be calculated.  Once the
gravitational potential is computed, it is then interpolated back to
the particles.  This process can lead to a loss of accuracy due to the
required interpolations.  It might also spuriously generate entropy if
the particle resolution is too low to adequately sample the density
field \cite{spri10}.  This might be the key to understand the
differences seen in the central entropy profiles of galaxy clusters
simulated with SPH and mesh-based Eulerian techniques \cite{mitc09}.
Eventually, the interpolation processes increase the communication
overhead in massively parallel simulations \cite{mvh13}.  A possible
remedy in grid-based codes is the stellar hydrodynamical approach
\cite{lars70, bh87}.  With this approach, the stars are treated as a
collisionless fluid and their evolution is regulated by the moments of
the Boltzmann equation.  This approach has been used many times to
simulate galaxies \cite{tbh92, sht97, vt06, vt08}.  Recently, Mitchell
et al.  \cite{mvh13} implemented this method into the FLASH code.
Numerical tests confirmed the validity of this approach and the
advantages over the more conventional particle schemes.

Another very important aspect of the evolution of galaxies is the
multi-fluid, multi-phase treatment.  Stars and gas exchange mass,
momentum and energy during the whole life of the stars.  Also dust and
gas exchange mass and momentum (see Sect. 6 for more detail on
dust-gas interactions).  Moreover, various gaseous phases are known to
exist in the ISM and phase transformations occur continuously during
the life of a galaxy.  Eventually, the gas in the ISM is composed of
many different elements, with various ionisation states.  A complete
treatment of the galaxy evolution must take into account the various
phases of a galaxy and all possible exchange processes among them.  In
the classical chemodynamical approach, put forward by Hensler and
collaborators \cite{bh87, hb90, tbh92} stars and various gas phases
(typically a cold and a warm-hot phase) co-exist within a single grid
and exchange mass, momentum and energy according to physically-based
recipes.  The dynamics of the various phases might or might not be the
same.  Typically, the various gas phases share the same velocity field
whereas the dynamics of the stars are different.  This approach has
been refined over the years and many groups use it to simulate
galaxies , with various degrees of sophistication \cite{sc02, cc02,
  htg04, hth06, scan06, mura10, ph11}.  Nowadays, chemodynamics is a
widely used term that generically refers to simulations in which some
treatment of the chemical evolution is included \cite{rmd01, kg03,
  koba04, torn04, few12}.  Although these codes clearly represent a
step forward with respect to more traditional single-fluid
simulations, still they lack the complexity of the multi-phase
chemodynamical codes described above.

Unfortunately, not many works in the literature have been devoted to
the direct comparison of single-phase and multi-phase models.  In
simulations of the hot interstellar medium in elliptical galaxies
\cite{fuji97}, the treatment of SNe ejecta as a separate phase makes
SN explosions less effective at heating the ISM because most of the
explosion energy is released in a dense and metal-rich medium and it
is quickly radiated away.  The SN energy is more efficiently
transformed to thermal energy of the ISM in single-phase simulations.
The multi-phase description of the ISM in simulations of ram-pressure
stripping (see Sect. 8) changes the distribution of gas compared to a
single-phase model \cite{tb09}.  This is due to the fact that the ISM
in multiphase simulations is more structured and with larger density
variations.  Steep density gradients are much better resolved in SPH
multi-phase implementations compared to classical single-phase ones
\cite{rt01}.  The multi-phase treatment can be the key to solve the
so-called overcooling problem typically encountered in single-phase
simulations (see Sect. 7 for more discussion on that aspect) .

Also the metallicities and abundance ratios of simulated galaxies can
be significantly affected by the multi-phase treatment of the ISM.
For instance, the presence of a cloudy phase dilutes the ISM, without
preventing the formation of large-scale outflows, able to eject a
fraction of the freshly produced heavy elements (see Sect. 9).  The
resulting final metallicity of model galaxies with a multiphase
(cloud-intercloud) ISM treatment is therefore generally lower (by
~0.2-0.4 dex) than the one attained by single-phase models
\cite{rh07}.  A clear example of the effect of a cloudy medium is
presented for instance in fig. 7 of \cite{rh07}.  Clouds (in
particular infalling clouds) produce not only a decrease in the global
metallicity, but also a variation in the C/O, N/O and Fe/O abundance
ratios.  In particular, the observed N/O abundance in the galaxy I Zw
18 seems to require the presence of infalling clouds.

\section*{4. The star formation}
In spite of still many open questions, enormous progresses have been
made in the last decade in simulating the process of star formation
\cite{bbb03, km05, bb06, krum09, bate12, greif11}.  However, the level
of detail and the resolution reached by these works can not be matched
by galactic simulations.  Suitable parametrisations of the star
formation need to be implemented.  It is also worth mentioning that
many papers dealing with simulations of galaxies do not
self-consistently calculate the star formation, but use prescribed
star formation rates (SFRs) or star formation histories (SFHs).  These
are either based on the reconstructed SFH of specific galaxies
\cite{recc04, recc06}, or are simple functions of time such as
instantaneous bursts or exponentially declining SFRs \cite{tosi88,
  mf99, recc02, frag04}.  This is a viable possibility if the star
formation process itself is not the focus of the numerical study.

A star formation law scaling with some power of the gas volume or
surface density is often assumed.  This relation is based on the
observation of star formation indicators in local galaxies
\cite{kenn98} and is often called the Kennicutt-Schmidt law.  To be
more precise, the Kennicutt-Schmidt law implies that:
\begin{equation}
\Sigma_{\rm SFR} \propto \Sigma_g^n,
\label{eq:ks}
\end{equation}
\noindent
where $\Sigma_{\rm SFR}$ is the SFR surface density and $\Sigma_g$ is
the gas surface density.  The value of $n$ reported by Kennicutt
\cite{kenn98} is 1.4$\pm$0.15.  In many works, a dependence on the
total volume density \cite{ta75, mg86, dr94, matt12} or on the
molecular gas density \cite{rw86, kmt09, krum13, hnm13} is also
assumed.  A dependence on the molecular gas density appears to be
particularly relevant because there is a tight correlation between the
H$_2$ and the SFR surface densities \cite{bigi08}.  Moreover, in spiral
galaxies, often the Toomre criterion is used to identify regions prone
to star formation \cite{vrh12}, or $\Sigma_{\rm SFR}$ is assumed to be
$\propto \Omega \Sigma_g$, where $\Omega$ is the circular frequency
\cite{ws89, pc99}.  Eventually, the spatial distribution of a
molecular cloud seems to play a critical role in determining its star
formation activity \cite{lada13}, but the dependence of the SFR on the
structure of a molecular cloud appears to be very difficult to
implement in numerical simulations.

In hydrodynamical simulations, many authors still follow the star
formation recipes of Katz \cite{katz92}, namely (see also Katz et al.
\cite{kwh96}):
\begin{itemize}
\item The gas density must be larger than a certain threshold
\item The particle must reside in an overdense region
\item The gas flow must be converging ($\nabla \cdot {\mathbf v} < 0$)
\item The gas particle must be Jeans unstable: $\frac{h}{c_s}>
  \frac{1}{\sqrt{4\pi G\rho}}$, where $h$ is the dimension of the gas
  particle (smoothing length for SPH simulations and the grid cell
  size for grid-based methods) and $c_s$ is the local sound speed
\end{itemize}
\noindent
With small variants, this recipe has been applied in most of galaxy
simulations \cite{wn99, slyz05, stin06, tb08, gove10}.  The Jeans
criterion appears to be particularly relevant, otherwise artificial
fragmentation and, hence, spurious star formation can arise
\cite{true97, bbb97}.  However, in some simulations the implementation
of this criterion has lead to unrealistic SFRs \cite{stin06}.

Often, a star formation law of the type:
\begin{equation}
\psi(t)=c_* \frac{\rho}{t_{\rm dyn}},
\label{eq:sflaw}
\end{equation}
\noindent
is assumed, where $\psi(t)$ is the SFR and $c_*$ is the star formation
efficiency \cite{stin06, scan06, delu06}.  Here $t_{\rm dyn}$ is a
typical star formation timescale given by the free-fall timescale, the
cooling timescale or a combination of both.  Notice that the free-fall
time scale is proportional to $\rho^{-1/2}$, thus a star formation
very similar to the Kennicutt-Schmidt law can be obtained in this way
(see also \cite{elme02}).  Notice also that observed laws (such as the
Kennicutt-Schmidt law Eq. \ref{eq:ks}) involve surface densities,
whereas theoretical models and simulations generally work with volume
density laws such as Eq. \ref{eq:sflaw} and not necessarily these two
formulations are equivalent.  Typically adopted values for $c_*$ in
Eq. \ref{eq:sflaw} are quite low, ranging between 0.1 and 0.01
\cite{stin06}.  This is also the ratio between the gas consumption
time scale and $t_{\rm dyn}$.  This assumption is in agreement with
the conclusion, deduced from observations, that only a small fraction
of gas in molecular clouds can be converted into stars \cite{evans09,
  murr11}.  The star formation efficiencies are larger (of the order
of 0.3) if one considers only the dense cores of molecular clouds
\cite{all07}.  Global star formation efficiencies tend to be even
lower in DGs (see also Sect. 2 and below in this Section).

As mentioned above, the fraction of molecular gas is taken into
account in some star formation recipes.  In particular, the right-hand
side of Eq. \ref{eq:sflaw} is often multiplied by $f_{\rm H_2}$, i.e.
by the ${\rm H}_2$ mass fraction \cite{kmt09, krum13}.  It has been
shown \cite{kuhl12, kuhl13} that such a star formation law applied to
cosmological simulations leads to a strong reduction of the star
formation in low-mass halos compared to models without molecular
fraction dependencies.  This might help explain the mismatch between
the observed mass distribution of DGs and the predictions of the
$\Lambda$-CDM theories (see also Sect. 2).

Since the cooling timescale depends on the gas temperature, a
dependence of the star formation with the temperature is implicit in
Eq. \ref{eq:sflaw}.  It is of course very reasonable to assume that
the SFR depends on the temperature, since star formation occurs in the
very cold cores of molecular clouds.  For this reason, some authors
even assume a temperature threshold, above which star formation cannot
occur \cite{stin06, recc07, ager12}.  However, one should be aware of
the fact that simulations still do not have the capability to
spatially resolve the cores of molecular clouds.  The temperature of a
star forming region is thus simply the average temperature of a region
of gas, with size equal to a computational unit (gas particle in a SPH
simulation or grid cell in grid-based codes), encompassing the star
forming molecular cloud core.  For this reason, typical temperature
thresholds are of the order of 10$^3$--10$^4$ K, at least two orders
of magnitude larger than typical molecular core temperatures.  

\begin{figure*}[t]
\resizebox{10.cm}{!}{\includegraphics{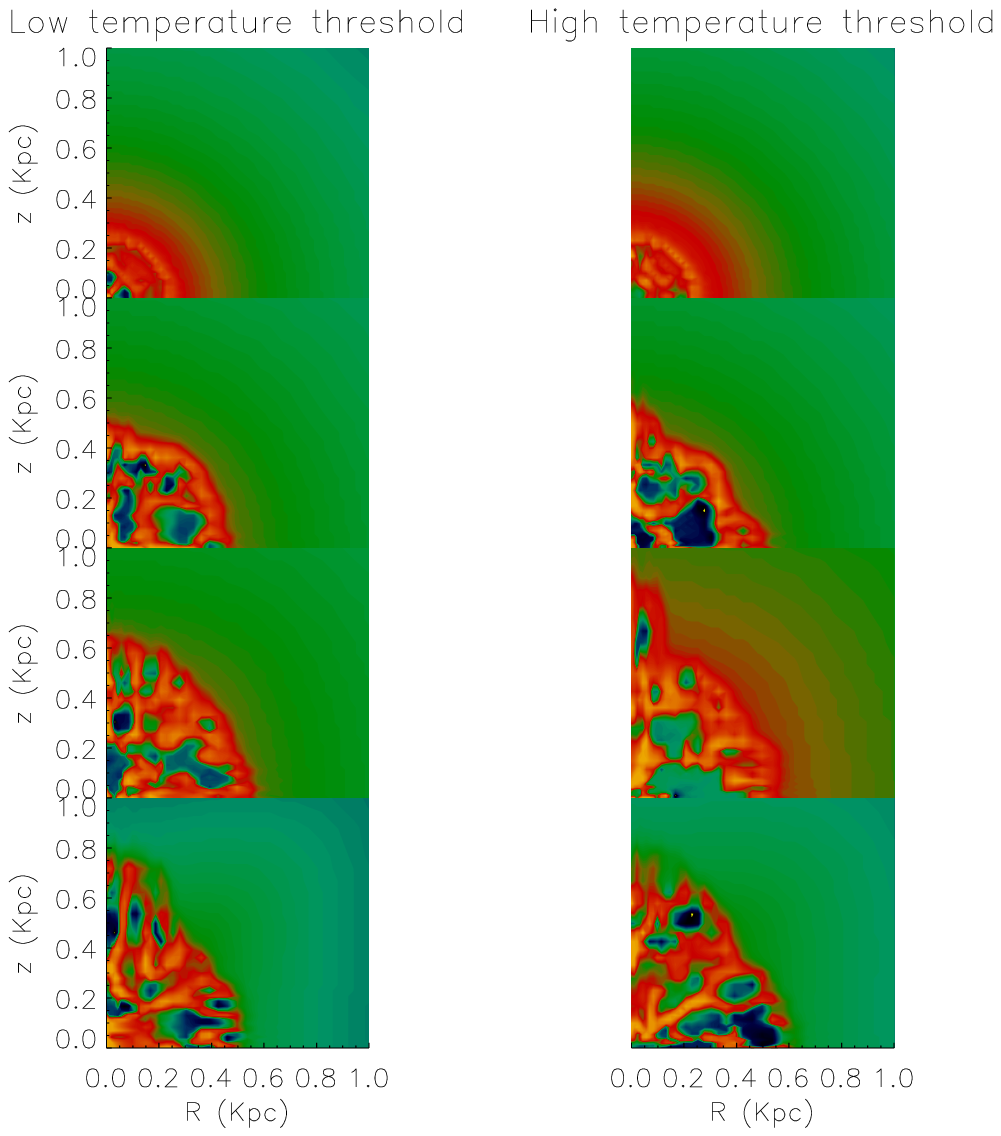}}
\resizebox{7.cm}{!}{\includegraphics{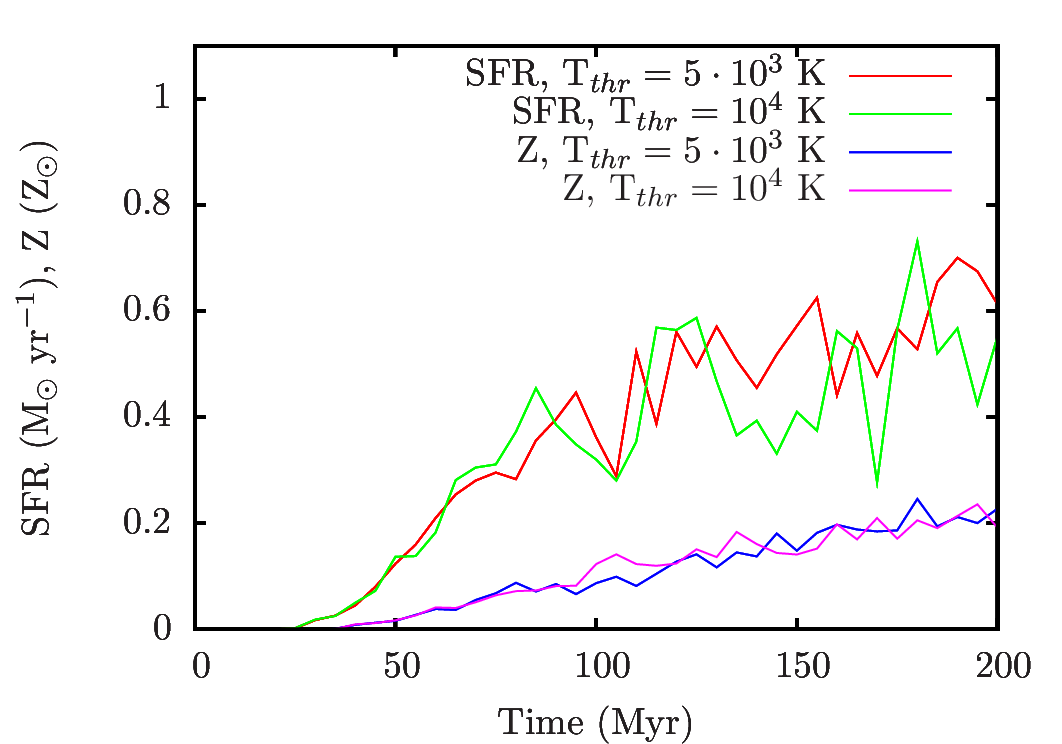}}
\caption{The effect of the temperature threshold T$_{thr}$ above which
  star formation is not allowed. {\bf left)} The density distribution
  of gas in two runs with different T$_{thr}$: T$_{thr}=5\cdot 10^3$ K
  (left panels) and T$_{thr}=10^4$ K (right panels).  The four rows
  are snapshots of the evolution of the two models at four different
  moments in time: 50 Myr (upper-most panels), 100 Myr (second row of
  panels), 150 Myr (third row of panels) and 200 Myr (lower-most
  panels).  Dense gas is in orange (upper densities are 10$^{-23}$ g
  cm$^{-3}$); dilute gas in blue (lower densities are 10$^{-27}$ g
  cm$^{-3}$).  {\bf right)} Star formation rates (in M$_\odot$
  yr$^{-1}$) and global metallicities (in Z$_\odot$) for the same
  models shown in the left panels, during the first 200 Myr of
  galactic evolution.}
\label{fig:tthrdep}
\end{figure*}

Some authors adopt a more complex temperature dependence.  For
instance, K\"oppen et al.  \cite{kth95} derive:
\begin{equation}
\psi(t)=c_* \rho^2 e^{-T/T_s},
\label{eq:kth95}
\end{equation}
\noindent
where the transition temperature $T_s=1000$ K implies that the star
formation is very low in regions with $T>T_s$.  Notice that, in this
case, $c_*$ does not have the same dimensions (and the same meaning)
of the $c_*$ introduced in Eq. \ref{eq:sflaw}.  This star formation
recipe, coupled with the feedback from stellar winds and dying stars
(see Sect. 7), nicely leads to self-regulation of the star formation
process.  In fact, a large SFR increases the feedback, which in turn
strongly reduces further star formation whereas, if the feedback is
low, the temperature does not increase and star formation is more
efficient.  Because of the self-regulation, the star formation process
is not very dependent on the adopted parameters $c_*$ and $T_s$.

Eventually, theoretical works \cite{ee97} suggest that the star
formation efficiency can depend on the external pressure, simply
because gas collapse is favoured in environments with large pressures.
This hypothesis is supported by the observational fact that the
molecular fraction depends on the gas pressure \cite{br06, leroy08}
and, as noticed above, the surface density of molecular gas strongly
correlates with the SFR \cite{bigi08}.  DGs are usually characterised
by lower pressures compared to larger galaxies, thus the predicted
star formation efficiency is lower.  This finding is in agreement with
other lines of evidence, showing that DGs are quite inefficient in
forming stars (see Skillman et al.  \cite{skil12} for a review; see
also Sect. 2).  The pressure dependence on the star formation
efficiency has been used in Harfst et al. \cite{hth06}.

Various works in the literature have been devoted to the comparison of
different star formation schemes in simulations of galaxies.  I will
briefly summarise some of these works, but before doing so, it is
important to remark what written above: the star formation process (if
adequately simulated) tends to self-regulate, therefore moderate
variations of the involved parameters produce little changes in the
final outcomes of the simulations.  Fig.  \ref{fig:tthrdep} shows an
example of this self-regulation.  The outcomes of models simulating
DGs without massive dark matter halos (modelled as in \cite{recc07})
are shown.  Two values of the temperature threshold $T_{thr}$, above
which the star formation is not allowed, have been adopted.  As one
can see, the evolution of these two model galaxies (distribution of
gas, SFR, evolution of the global metallicity) is fairly insensitive
to the chosen value of $T_{thr}$.

A thorough investigation of different prescriptions for turning cold
gas into stars in SPH cosmological simulations \cite{kay02} shows that
the results are indeed fairly insensitive to many parameters
describing the star formation process (temperature and density
thresholds, overdensity threshold).  However, large differences in
these parameters might indeed lead to completely different results
(see e.g.  \cite{gove10} for differences in DG models with density
thresholds varying by up to four orders of magnitude).  Also relaxing
the criterion $\nabla \cdot {\mathbf v} < 0$ seems not to produce
large differences in some simulations of galaxies \cite{buon00}.
Variations of the parameter $c_*$ might instead lead to different
results, at least in some implementations. The average SFR becomes
larger for large values of $c_*$ and a good fit with the observed
Kennicutt-Schmidt law is obtained with $c_*=0.05$ \cite{stin06}.

\section*{5. The initial mass function}
Once the stars are born, a mass distribution must be assumed.  In
fact, the chemical and mechanical feedback of massive stars
substantially differ from the feedback of low-and intermediate-mass
stars (see next sections), thus it is crucial to know how many stars
are formed per each mass bin.  Actually, the IMF is often combined
with the SFR to obtain the so-called birthrate function $B(m,t)$
\cite{tins80, matt12}, which gives the number of stars formed per unit
stellar mass and per unit time.  Usually, the time dependence is
described by the SFR, whereas the mass dependence is determined by the
IMF.  However, one should already point out that, according to some
lines of evidence, the IMF could depend on time, too (see below).

The IMF $\xi(m)$ was originally defined by Salpeter \cite{salp55} as
the number of stars per unit logarithmic mass that have formed within
a specific stellar system.  Thus, the total mass of stars with masses
between $m$ and $m+dm$ is $\xi(m)dm$.  A very useful concept is also
the IMF in number $\varphi(m)$, giving the number of stars in the
interval [$m$, $m+dm$].  Clearly, $\xi(m)=m\varphi(m)$.  Salpeter
found out that $\xi(m)\propto m^{-1.35}$ for 0.4 M$_\odot$ $<m<$ 10
M$_\odot$.  This estimate has been refined over the years
\cite{tins80, scalo86, ktg93, chab03} and nowadays a commonly used
parametrisation is the so-called Kroupa IMF \cite{krou01}, namely a
three-part power law $\xi(m)\propto m^{-\gamma}$ with $\gamma=-0.7$ in
the interval 0.01 M$_\odot$ $<m<$ 0.08 M$_\odot$ (i.e. in the brown
dwarf domain), $\gamma=0.3$ for 0.08 M$_\odot$ $<m<$ 0.5 M$_\odot$,
and finally $\gamma=1.3$ (very similar to the Salpeter slope) for
stellar masses larger than 0.5 M$_\odot$.  

The paper of Romano et al.  \cite{roma05} clearly shows how different
IMFs can change the fraction of stars in various mass bins (see their
table 1).  IMFs predicting smaller fractions of massive stars produce
less $\alpha$-elements, because these elements are mainly synthesised
by SNeII.  This is evident in fig. 6 of \cite{roma05}, which shows the
evolution of [$\alpha$/Fe] vs.  [Fe/H] for model galaxies
characterised by different IMFs.  Since more massive stars means more
SNeII, clearly the IMF affects the energetics of a galaxy, too.  This
has been shown in many simulations \cite{torn04, scha10, tesc11,
  wire11}.  In particular, flat IMFs tend to produce higher fractions
of massive stars and, hence, larger SNeII luminosities.  The energy
supplied by SNeII could be enough to unbind a fraction of the ISM and
produce a galactic wind (see also Sect.  9).

It is important to point out that, usually, numerical simulations
adopt a fixed value for the IMF upper stellar mass $m_{\rm up}$,
irrespective of how much gas has been converted into stars.  However,
$m_{\rm up}$ should depend on the mass of the newly formed stellar
particles, for the simple reason that only massive star clusters can
host very massive stars.  A correlation between the stellar cluster
mass $M_{\rm cl}$ and the upper stellar mass is indeed observationally
established and can be reproduced by simply assuming that $m_{\rm up}$
is the mass for which the IMF in number $\varphi(m)$ is equal to 1
\cite{kw03}.  Weidner \& Kroupa \cite{wk06} found that the
theoretically derived $M_{\rm cl}$-$m_{\rm up}$ relation nicely
reproduces the available observations (their figs. 7 and 8; see also
\cite{wkp13}).  Clearly, this assumption can greatly affect the
outcomes of simulations, but, to the best of my knowledge, it has
never been explored in detail in hydrodynamical simulations of
galaxies.

\begin{figure*}[t]
\resizebox{8.5cm}{!}{\includegraphics{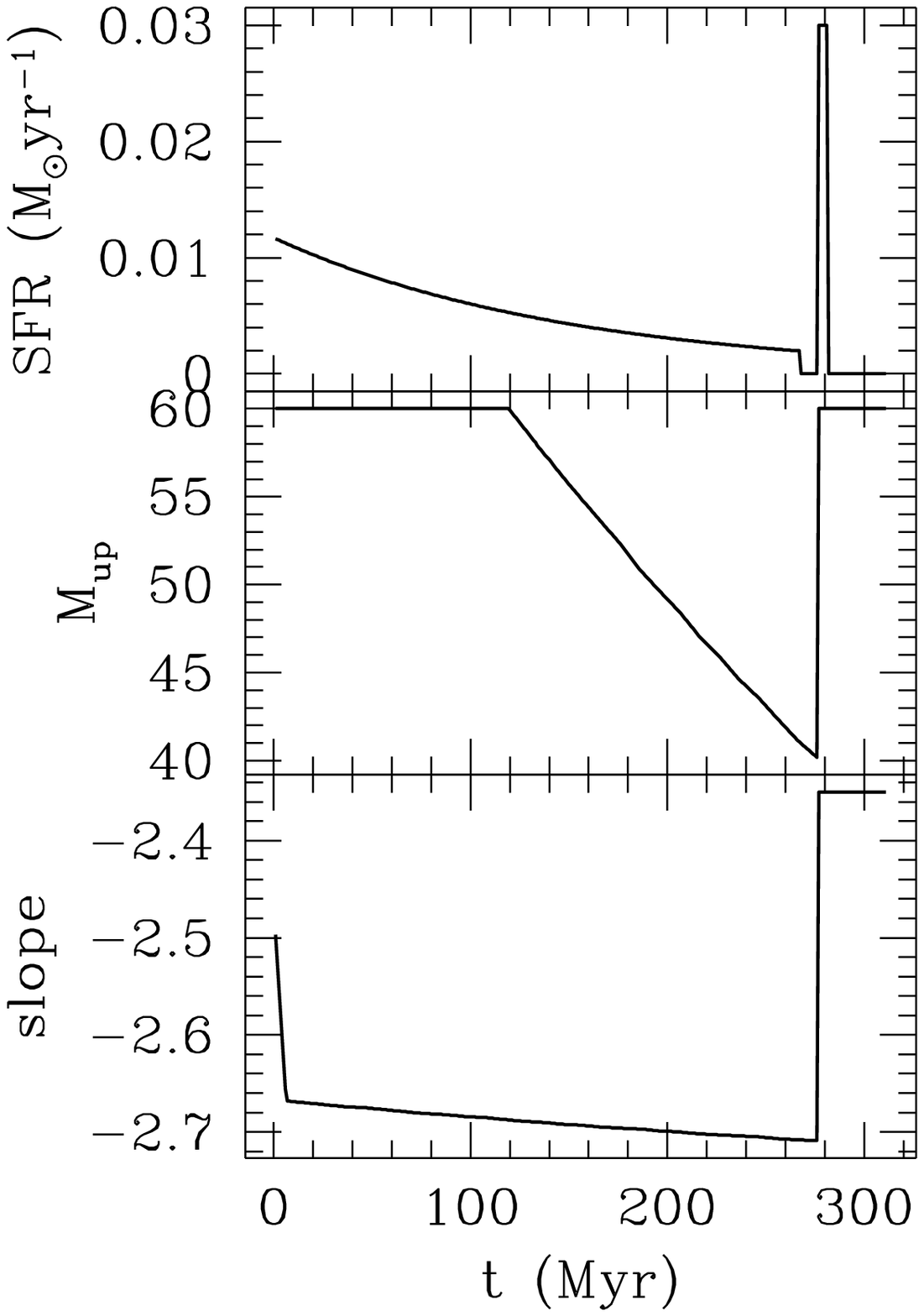}}
\resizebox{8.5cm}{!}{\includegraphics{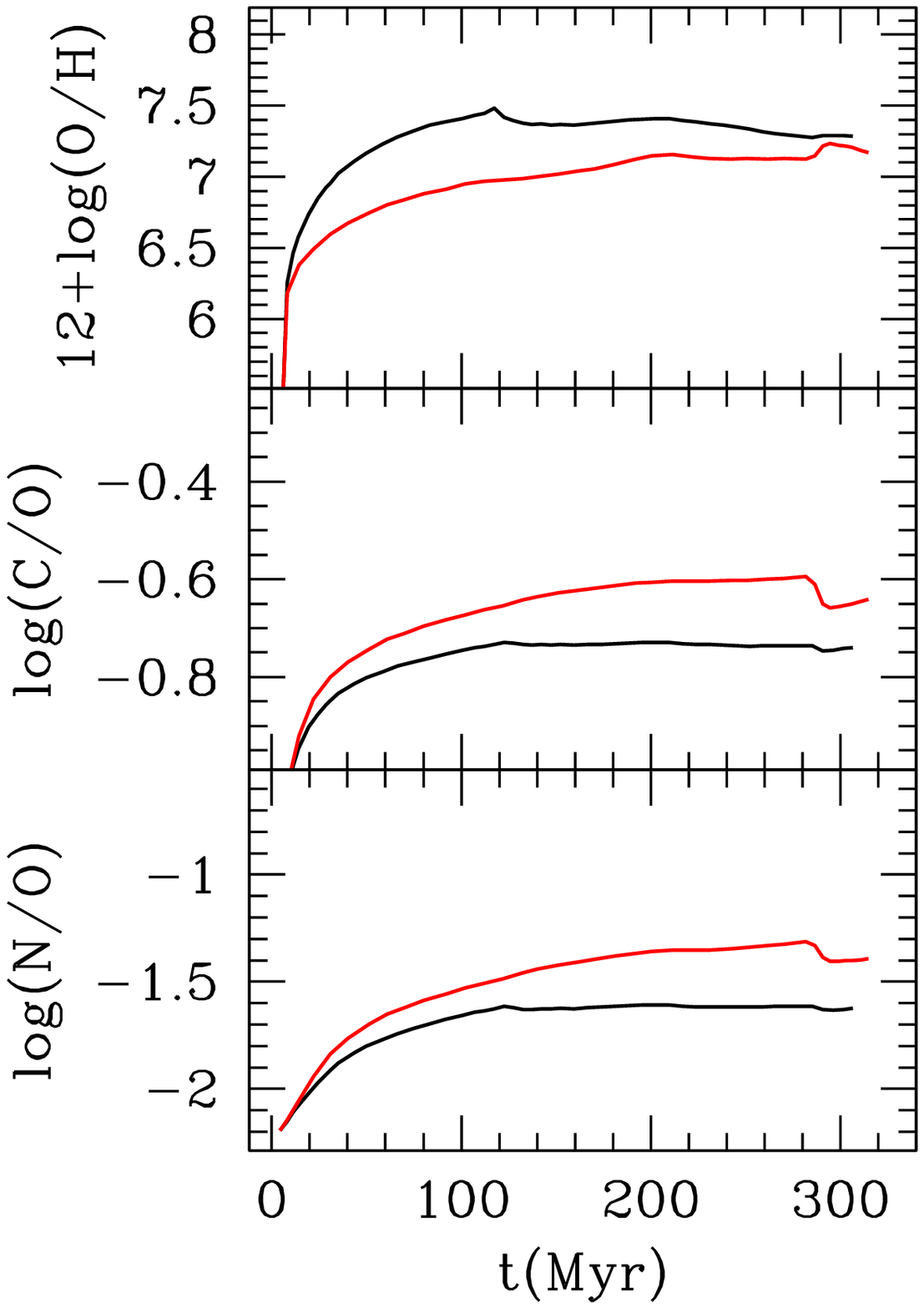}}
\caption{The effect of the IMF on the evolution of galaxies. {\bf left)} 
  The adopted SFR (upper panel), together with the upper stellar mass
  $M_{\rm up}$ (in M$_\odot$, middle panel) and the average slope of
  the IMF (in number, lower panel) calculated for the IGIMF galactic
  model (red lines in the right panels).  {\bf right)} Predicted
  evolution of abundances and abundance ratios for a IGIMF galactic
  model (red lines).  Plotted are the evolution of oxygen (upper
  panel), carbon-to-oxygen ratio (middle panel) and nitrogen-to-oxygen
  ratio (lower panel).  The black line represents the evolution of a
  model with a time-independent Salpeter IMF (i.e. with a slope of
  -2.35).}
\label{fig:igimf}
\end{figure*}

Since a correlation between the most massive cluster in a galaxy and
the SFR $\psi$ is also observationally established \cite{wkl04}, the
logical consequence is that the galaxy-wide IMF in a galaxy must
depend on the SFR, too.  In particular, the IMF is time-dependent and
is given by the integral of the IMFs of single star cluster, which are
assumed to always be a Kroupa IMF, but with different upper masses
$m_{\rm up}$, depending on the star cluster mass.  An upper cluster
mass limit depending on $\psi$ is then assumed.  Given a mass
distribution of embedded clusters $\varphi_{\rm cl}(M_{\rm cl})$
(giving the number of star clusters in the interval [$M_{\rm cl}$,
$M_{\rm cl}+dM_{\rm cl}$]), the global, galactic-scale IMF (integrated
galactic IMF or IGIMF) is given by:
\begin{equation}
\varphi_{\rm IGIMF}=\int_{M_{\rm cl, inf}}^{M_{\rm cl, sup}(\psi)} 
\varphi\left(m<m_{\rm up}(M_{\rm cl})\right)\varphi_{\rm cl}(M_{\rm cl})
dM_{\rm cl},
\label{eq:igimf}
\end{equation}
(see \cite{kw03, wk05, rck09} for details.  Notice also that in the
original papers the IMF in number is designed with $\xi$ instead of
with $\varphi$).  The IGIMF turns out to be steeper than the Kroupa
IMF assumed in each star cluster and the difference is particularly
significant for low values of the SFR.  Notice however that the IMF
tends to become top-heavy when the SFR is very high \cite{wkp11}.  The
effect of the IGIMF on the chemical evolution of galaxies has been
already explored in a few papers \cite{kwk07, rck09, calu10, recc13}.
It turns out that the IGIMF is a viable explanation of the low
metallicity \cite{kwk07} or of the low $\alpha$/Fe ratios \cite{rck09}
observed in DGs.  The main reason is that DGs have on average lower
SFRs and this, in turn, implies steeper IMFs, characterised by a lower
fraction of massive stars.  The production of metals and, in
particular, of $\alpha$-elements, is considerably reduced.

Chemo-dynamical simulations of galaxies can give a more complete
picture of the evolution of DGs and of the effect of the IMF (and of
the IGIMF, in particular).  Fig. \ref{fig:igimf} shows the comparison
of the results of two chemo-dynamical simulations, with and without
adopting the IGIMF.  Methods, assumptions and initial conditions are
taken from \cite{recc04}.  In particular, the main structural
properties of the shown model galaxies resemble the blue compact DG
IZw 18 (see \cite{vi98, po12} for a summary of observed properties of
this galaxy).  The SFH is shown in the upper left panel.  This
particular dependence of the SFR with time has been chosen again in
agreement with the reconstructed SFH of IZw 18 as derived by
\cite{atg99} (but see \cite{anni13} for a more recent determination of
the SFH in IZw 18).  According to this SFH, the IGIMF predicts
variations of the upper stellar mass and of the average IMF slope as
shown in the middle and lower panels, respectively.

The evolution of gas-phase abundances and abundance ratios in a
simulation adopting these IGIMF prescriptions is shown in the right
panels (red lines) and compared with the results obtained with a model
adopting a standard, time-independent Salpeter IMF (black lines).
Since the IGIMF is steeper (and poorer in massive stars) than the
Salpeter IMF, the initial phases are characterised by a lower
production of oxygen and, consequently, higher values of C/O and N/O.
However, due to the higher feedback, the model with Salpeter IMF
experiences a galactic wind at $t\simeq 120$ Myr.  Since galactic
winds tend to be metal-enriched (see also Sect. 9), the onset of the
galactic wind is characterised by a decrease in O/H.  The galactic
wind does not occur in the IGIMF run due to the reduced number of
SNeII.  At $t\simeq 280$ Myr a burst of star formation occurs (see
upper left panel).  In the Salpeter IMF run, most of the freshly
produced metals are channelled out of the galaxy and do not contribute
to the chemical enrichment.  In the IGIMF run instead, the metals
newly synthesised during the burst do contribute to the chemical
enrichment and this causes a sudden increase of the oxygen abundance
(and a sudden decrease of C/O and N/O).  

\begin{figure*}[t]
\resizebox{8.5cm}{!}{\includegraphics{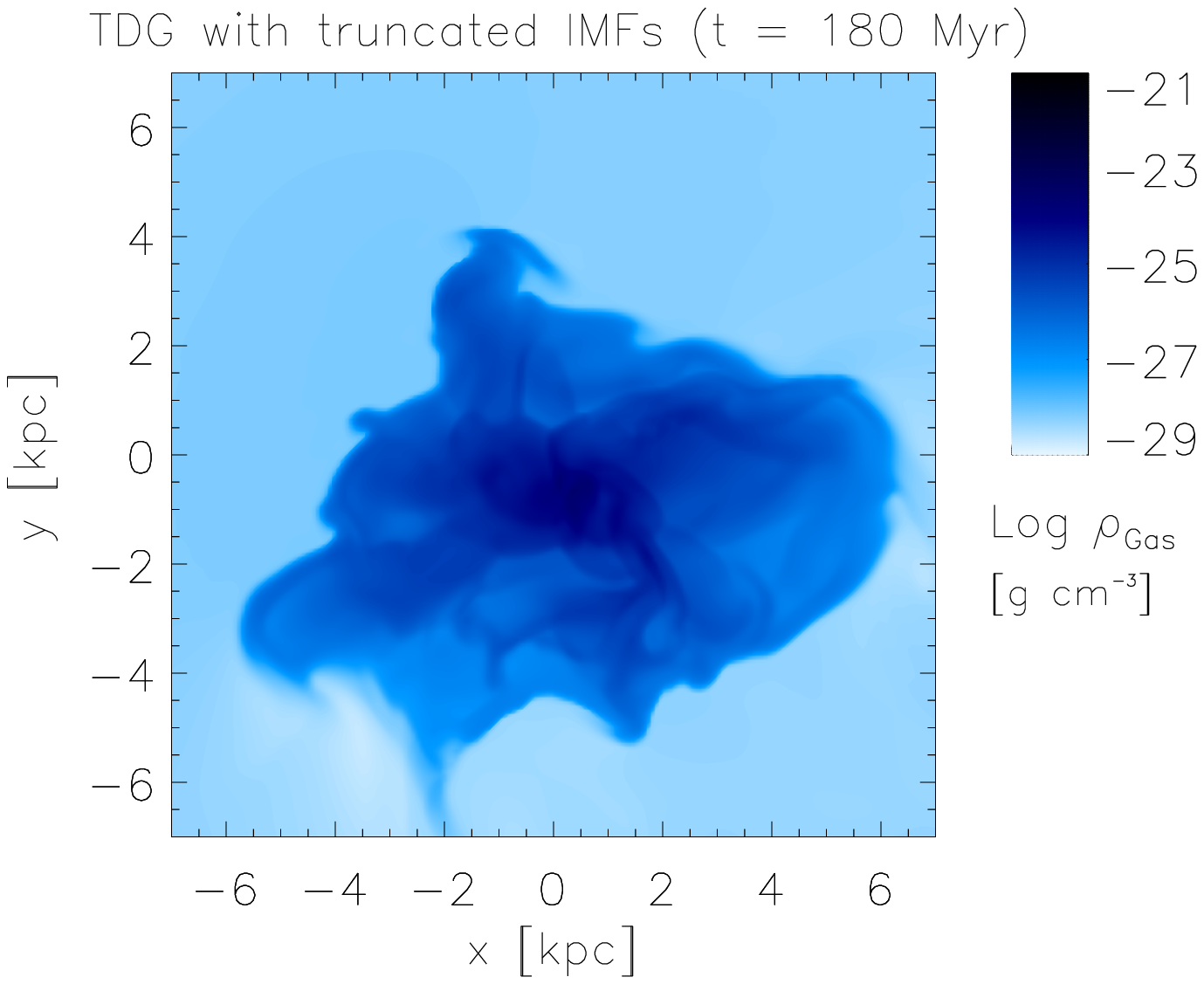}}
\resizebox{8.5cm}{!}{\includegraphics{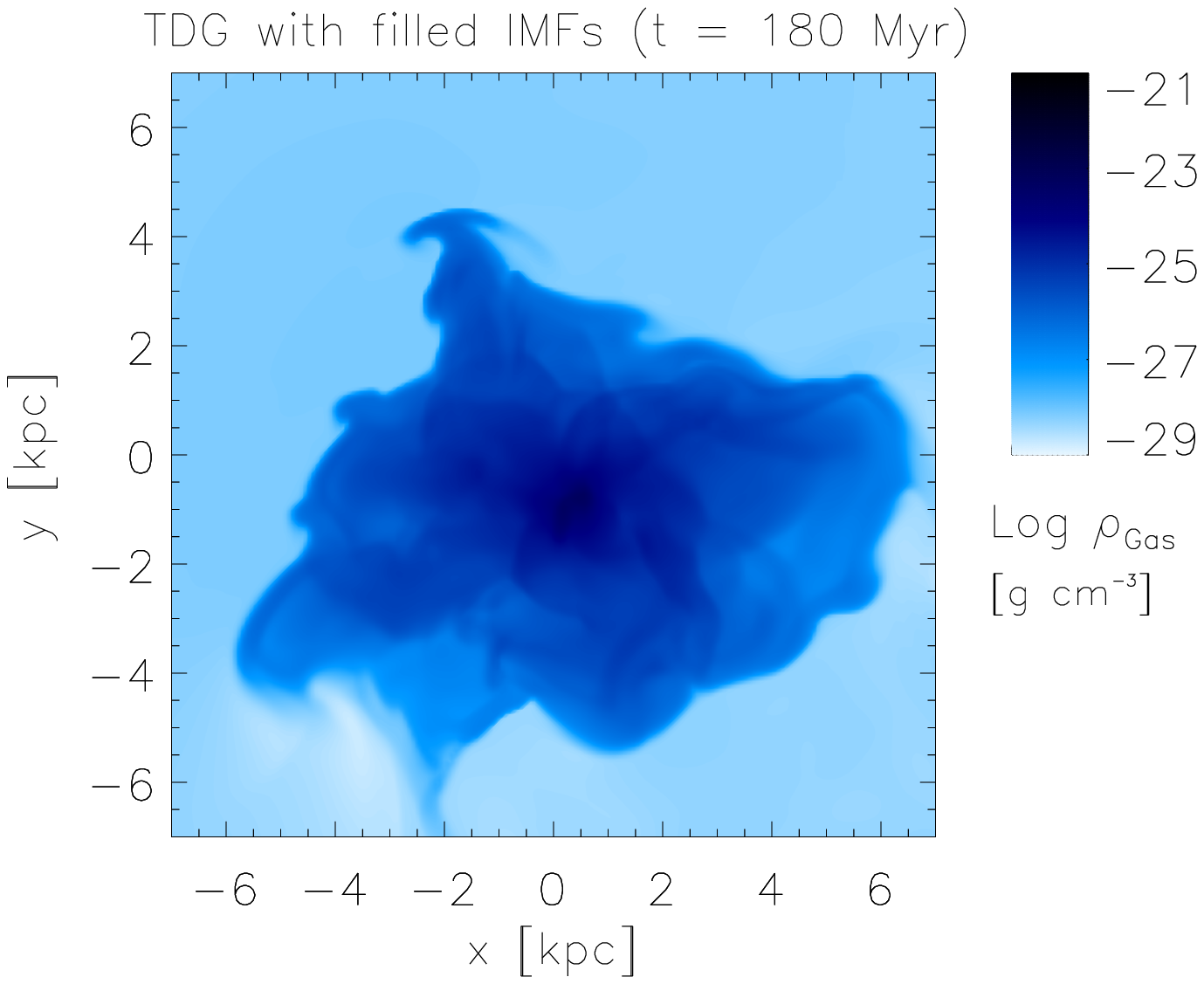}}
\resizebox{8.5cm}{!}{\includegraphics{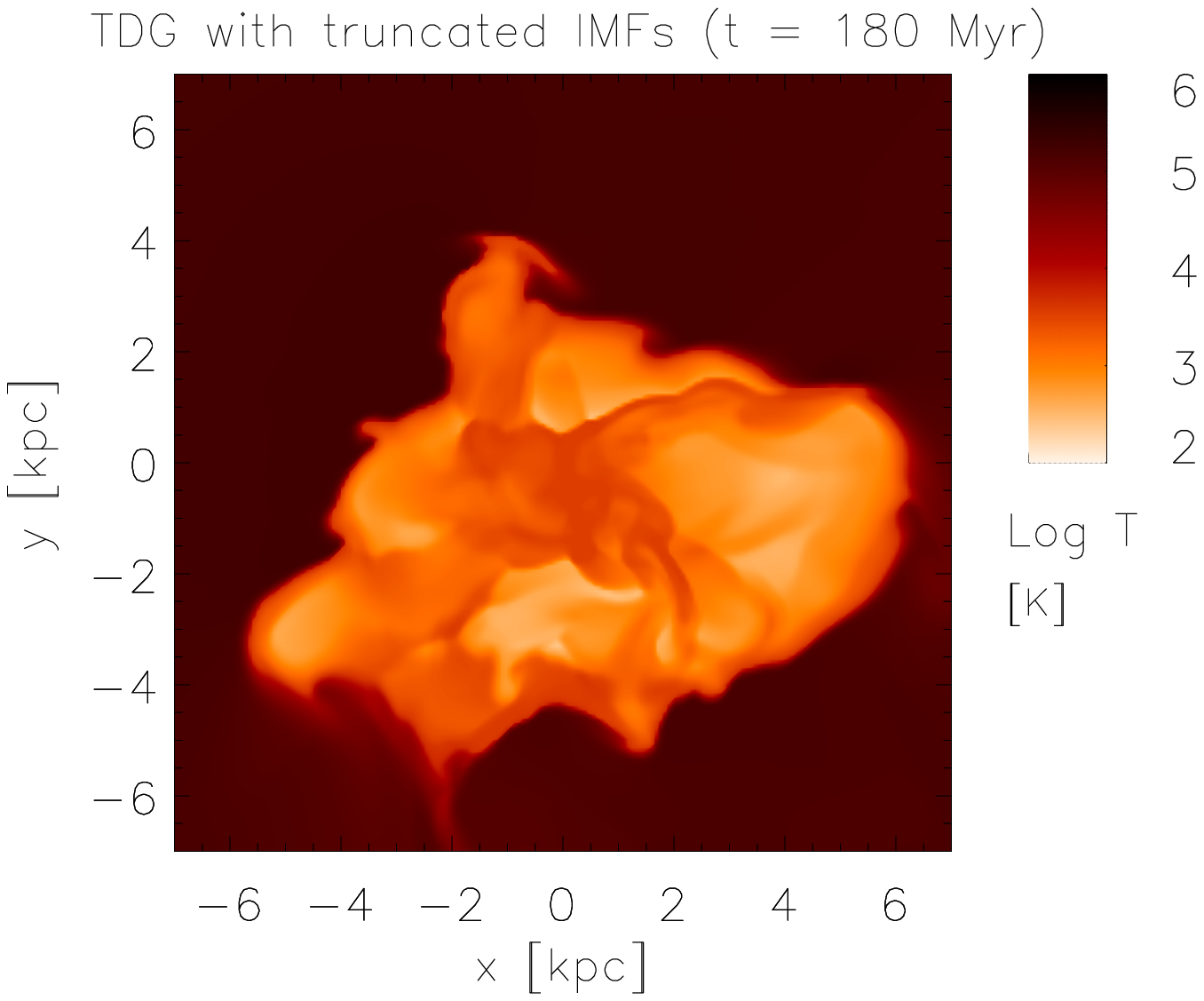}}
\resizebox{8.5cm}{!}{\includegraphics{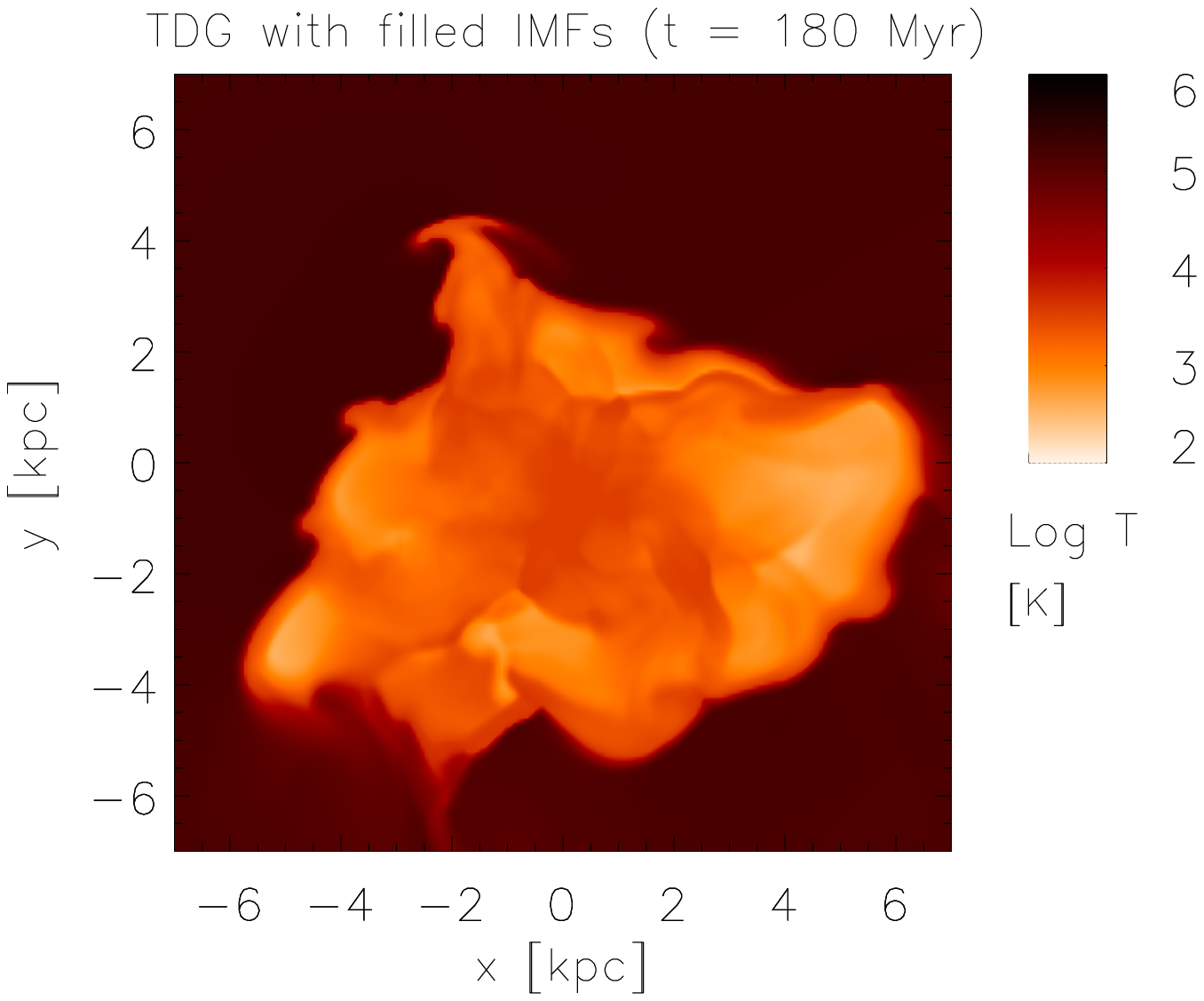}}
\caption{The density (upper panels) and temperature (lower panels)
  distributions after 180 Myr of evolution of two model galaxies.
  Strips on the right of each panel show the density and temperature
  scales.  In one case (left panels), the upper mass in each formed
  star cluster depends on the total cluster mass (truncated IMF).  In
  the other case (right panels), the upper IMF mass is always the
  same, irrespective of the mass of the star cluster.  It is to notice
  that the temperature of the gas outside the galaxy has been set to
  10$^6$ K.  The temperature in the central region is hotter in the
  filled IMF model because of the larger amount of energy provided by
  SNeII.}
\label{fig:sylvia}
\end{figure*}

More detailed simulations, exploring wider parameter spaces, can show
other effects of the IGIMF.  In particular, the simulations shown in
Fig. \ref{fig:igimf} assume a pre-defined SFH, but it is clear that
the adoption of the IGIMF can affect the onset of the star formation,
too, because it affects the energetics of the ISM.  Numerical
simulations of galaxies with IGIMF and with star formation recipes as
described in Sect. 4 would surely predict different SFHs as compared
with models with SFR-independent IMFs.  This has been shown already in
chemical evolution models \cite{calu10} but this effect can be even
more dramatic in chemo-dynamical simulations.

An example of the effect of different IMF assumptions on the evolution
of galaxies is provided by Pl\"ockinger et al. \cite{ploe13}.  In this
paper, the star formation has been self-consistently modelled using
Eq.  \ref{eq:kth95}.  Stars organise themselves in clusters, whose
masses depend on the local reservoir of gas.  Within each cluster, it
is assumed that the mass of the most massive star correlates with the
total cluster mass, in compliance with the $M_{\rm cl}$-$m_{\rm up}$
relation described above (truncated IMF simulation).  This truncated
IMF model has been compared with a simulation in which the upper
stellar mass in each cluster is always the same, irrespective of the
cluster mass (filled IMF simulation).  The assumption of a truncated
IMF is particularly relevant for small clusters (with masses less than
a few 10$^2$ M$_\odot$): in these clusters, the most massive star is
smaller than 10 M$_\odot$, thus there are no (or very few) SNeII.
Since SNeII dominate the energy feedback in DGs (see Sect.  7), the
absence of these SNe lead to smaller temperatures compared to filled
IMF simulation.  This is shown in Fig. \ref{fig:sylvia}: the gas
density and temperature distributions of the truncated and filled IMF
simulations are compared at an evolutionary time of 180 Myr (when the
SFR reaches its peak).  The temperature in the truncated IMF model is
on average lower, thus the star formation can proceed for a longer
time and at a higher rate compared to the filled IMF simulation (see
also fig. 8 of \cite{ploe13}).

It is also important to point out that, in Eq. \ref{eq:igimf}, only
the global, galactic-scale SFR is required to calculate the IGIMF.
However, the star formation process is usually very inhomogeneous
within a galaxy, with regions of very enhanced star formation.
Clearly, the formation of massive stars is more likely in regions of
high star formation density.  It is reasonable thus to expect that the
IMF varies not only with time, but also with location within a galaxy.
This approach has been used for instance by Pflamm-Altenburg et al.
\cite{pk08} to explain the cut-off in H$\alpha$ radiation in the
external regions of spiral galaxies (where the SFRs are milder).  In
\cite{ploe13} also this effect can be appreciated (see in particular
their fig. 11).  Observational evidence of the variation of the IMF
within galaxies is given by Dutton et al. \cite{dutt13}.  To finish,
several lines of evidence point towards a dependence of the IMF on the
metallicity, too \cite{marks12, krou13}, in the sense that the IMF
appears to become top-heavy in metal-poor environments.  Clearly, the
chemo-dynamical simulations of galaxies with spatially and temporally
variable IMFs can give us new, different perspectives and insights to
understand the evolution of galaxies.

\section*{6. The chemical feedback}
In order to follow the chemical evolution of a galaxy, it is without
any doubt important to know how stars with different masses enrich the
ISM with various chemical elements.  The term stellar yields is
commonly used to indicate the masses of fresh elements produced and
ejected by a star of initial mass $m$ and metallicity $Z$.  However,
the term yields was originally introduced to indicate the ratio
between the mass of a specific chemical element newly created and
ejected by a stellar generation and the mass locked up in remnants
(brown dwarfs, white dwarfs, neutron stars and black holes; see also
Sect. 9).

\begin{figure*}[t]
\resizebox{17cm}{!}{\includegraphics{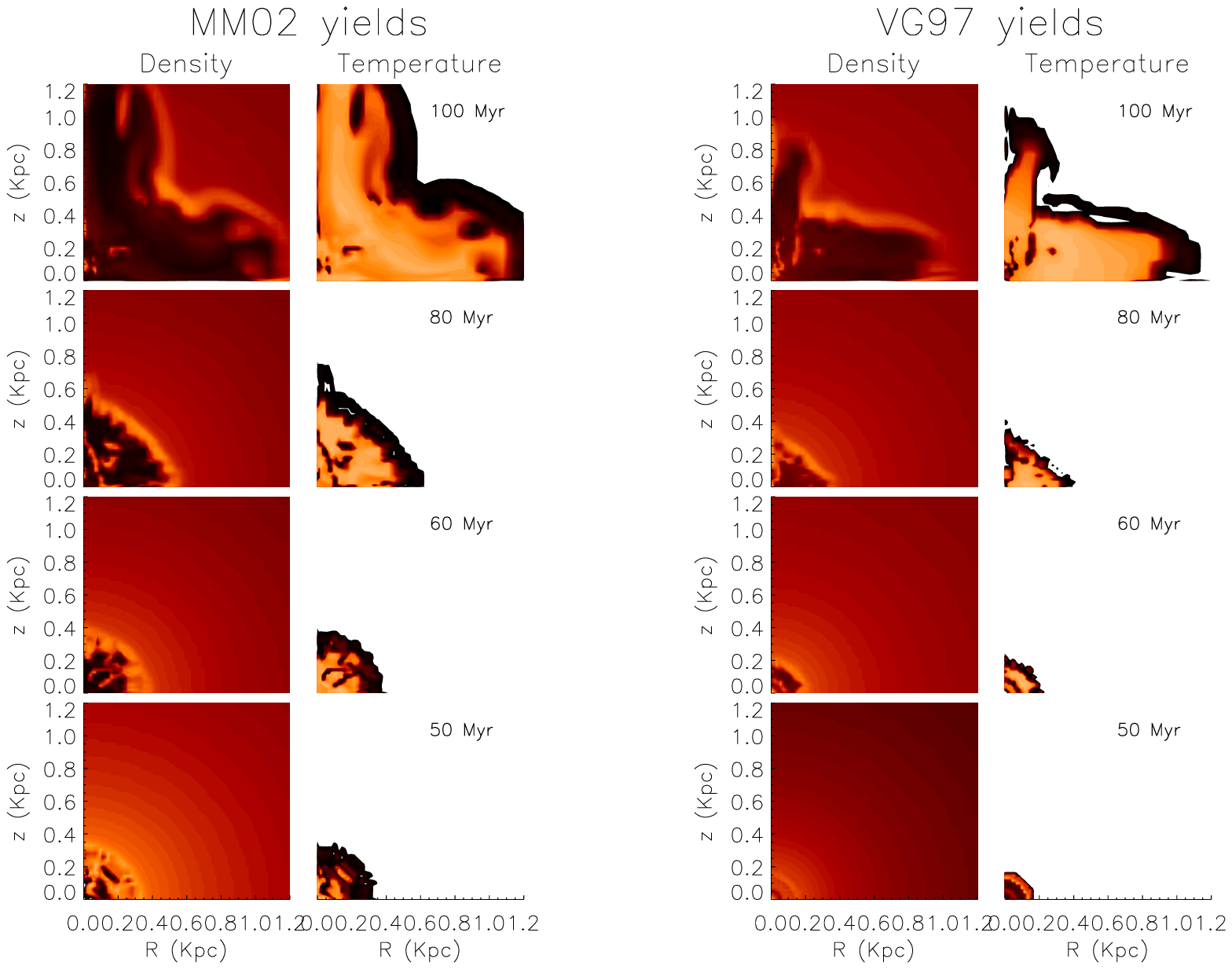}}
\caption{Density and temperature contours at 4 evolutionary times
  (labelled on each of the right panels) for a model adopting MM02
  (left) and VG97 (right) yields, respectively. The (logarithmic)
  density scale (in g cm$^{-3}$) ranges between -27 (dark) and -23
  (bright).  The (logarithmic) temperature scale (in K) ranges between
  3 (white) and 7 (orange).}
\label{fig:diffcool}
\end{figure*}

Many groups in the past few decades calculated the stellar yields of
both massive and intermediate-mass stars for different metallicities
\cite{rv81, ww95, vdhg97, pcb98, mm02, lc03, lc06, koba06, nkt13}.
Unfortunately, except for a handful of elements whose nucleosynthesis
in stars is well understood, yields of other elements calculated by
different authors can vary by orders of magnitude.  This is especially
true for the majority of the iron-peak elements, but also for much
more abundant species such as carbon and nitrogen (see the review of
Nomoto et al. \cite{nkt13}).  Of course, model predictions are
significantly affected by the choice of the set of yields.  This has
been shown by Romano et al. \cite{roma10} by means of neat and clear
numerical tests (see their figs 3 and 15, for instance).  One of the
most significant sources of uncertainty in the calculation of stellar
yields is the presence of stellar mass loss.  Massive stars with solar
metallicity might in fact lose a large amount of matter rich of He and
C, thus subtracting those elements to further processing, which would
eventually lead to the production of oxygen and other heavy elements.
The models of Maeder \cite{maed92} for instance predict that a 40
M$_\odot$ star ejects only $\sim$ 2 M$_\odot$ of O, whereas in most of
nucleosynthetic calculations without winds \cite{ww95, raus02, lc03}
the production of oxygen is a factor of $\sim$ 3 larger.

The yields from dying stars not only directly affect the chemical
composition of the ISM in chemo-dynamical evolution of galaxies, but
can also affect the dynamics by means of chemical feedback.  The main
effect is due to cooling.  In fact, it is known that the cooling
function of an optically thin plasma has a strong dependence on
metallicity, at least in the temperature range between $\sim$ 10$^4$
and 10$^5$ K \cite{bh89, sd93, schu09}.  Moreover, different chemical
elements contribute differently to the plasma radiative emission.
Clearly, the assumption of different yields in chemo-dynamical models
affects the chemical composition of the ISM, which in turn changes the
cooling timescales.  An example of the effect of different sets of
yields on the dynamical evolution of galaxies is given in Fig.
\ref{fig:diffcool}.  Two models of galaxy evolution (taken from the
suite of simulations of Recchi et al. \cite{recc07}) differ only in
the adopted nucleosynthetic prescriptions for intermediate-mass stars:
\cite{mm02} (MM02) on the left panels and \cite{vdhg97} (VG97) on the
right panels.  Yields of high-mass stars are in both cases taken from
\cite{ww95}.  Feedback from SNeII and stellar winds creates a network
of cavities and tunnels.  The superbubble evolution is faster in the
MM02 model.  Indeed, MM02 produces on average more metals, therefore
leading to larger cooling rates.  On the one hand, it reduces the
thermal energy content inside the superbubble, but on the other hand
this increased cooling favours the process of star formation, leading
to a more powerful feedback.  The latter effect prevails, and a larger
energy is available in model MM02 to drive the expansion of the
supershell.  Within the timespan of 100 Myr covered by these two
simulations, the differences between the two models are not huge.
They are, however, non-negligible and they tend to increase with time.
This simple test shows the effect of chemical feedback on the
evolution of a galaxy, an aspect that has been often overlooked in the
literature.

One should also be aware that other forms of chemical feedback operate
in galaxies.  The photoelectric emission from small dust grains and
PAHs can substantially contribute to the heating of the ISM
\cite{bt94}.  The amount of dust and PAH in a galaxy strongly
correlates with its metallicity \cite{lf98} and, consequently, the
metallicity affects the photoelectric heating of the gas.  It is
commonly assumed that for ISM metallicities below $Z_{\rm cr} \sim$
10$^{-5}$ Z$_\odot$, the star formation process is substantially
different and leads to a top-heavy IMF producing, on average, very
massive stars, the so-called PopIII stars \cite{schn02}.  As the ISM
metallicity approaches $Z_{\rm cr}$, the transition to a Salpeter-like
IMF occurs.

Under some circumstances, chemical reactions can affect the chemical
evolution, as well.  Astrochemistry is a vibrant and very active
astrophysical discipline \cite{dw84, tiel05} and nowadays the details
of many important atomic and molecular reactions occurring in the ISM
are known.  Although the chemistry of the dense gas in clouds is very
rich and variegate, less happens in the more dilute diffuse gas.
Global, galactic-scale simulations usually do not require the
implementation of complicated reaction networks.  However, the
presence of dust can significantly affect the chemical evolution.  It
is in fact well known that a large fraction of some chemical elements
(particularly Fe, Co, Ni, Ca, C and Si) is locked into dust grains
\cite{ss96}.  Clearly, it is impossible to have a complete picture of
the evolution of these chemical elements in the ISM without
considering the dust.  There have been several works about the
chemical evolution of galaxies with dust \cite{dwek98, zgt08, cpm08,
  pipi11, zh13}.  It is more complicated to include dust into
chemo-dynamical simulations of galaxies.  On the one hand, still not
much is known about the sources and composition of interstellar dust
\cite{tiel05}.  On the other hand, the physics of the dust-gas
coupling is still poorly known and typically assumed drag forces lead
to numerical problems \cite{mini10}.  In spite of these difficulties,
progresses have been made and simulations of galaxies taking into
account dust are becoming available \cite{to04, lp12}.  Clearly, this
is a field where more needs to be done.  Observations of dust in our
own Galaxy and in external galaxies are becoming extremely accurate
and the astronomical community is in dire need of detailed
chemo-dynamical simulations of dusty gases in order to help
interpreting the observations.

\section*{7. The mechanical feedback}

Explosions of SNe (of both Type Ia and II) and stellar winds are the
main drivers of the ISM dynamics, at least in DGs (in larger galaxies,
AGNs might play a fundamental role).  Unfortunately, for the
foreseeable future, galactic-scale simulations will not be able to
solve individual SN remnants or the effect of the wind from individual
stars.  Hence, heuristic, sub-grid recipes are needed to treat the
mechanical feedback.  This is a complex and still active research
field. Although feedback prescriptions have been found to address
specific issues \cite{gove10, dave10}, no recipe appears to be widely
applicable and physically justifiable.  Comparison studies have been
performed \cite{task08, scan12, kim13b}, and the overall conclusion
(see in particular the Aquila comparison project, \cite{scan12}) is
that the outcomes of numerical simulations crucially depend on the
feedback prescriptions and none of the considered codes is able to
satisfactorily reproduce the observed properties of the baryonic
component of galaxies.

Broadly speaking, feedback schemes can be divided into two categories:
kinetic feedback \cite{agui01, sh03, ds08} and thermal feedback
\cite{stin06, recc07, mona12, skory13}.  Kinetic feedback schemes are
mostly used in SPH simulations (but see \cite{dt08}).  The SN
explosion energy is transformed into kinetic energy of neighbouring
particles.  A kick is given to a few neighbouring particles, which
move after the kick with a prescribed velocity, along a random
direction.  The problem with this scheme is that it is not physically
justifiable and it is not easy to create galactic winds, unless kick
velocities are chosen along prescribed directions.

In thermal feedback schemes instead, the SN energy is used up to heat
the ISM. A well-known drawback of this scheme is that the cooling
timescale of the particles affected by this thermal feedback is
typically very short (often shorter than the timesteps of the
simulation).  The input energy is thus radiated away before it can be
converted to kinetic energy.  This leads to the so-called overvcooling
problem \cite{kwh96}.  Various authors have tried to remedy to this
problem by simply switching off the cooling \cite{sgp03, stin06,
  recc07}.  The inefficiency of thermal feedback is usually attributed
to poor spatial resolution: the energy is deposited in gas that is too
dense, because the hot, low-density, bubbles that fill much of the
volume of the multi-phase ISM are missing.  In fact, in models in
which the multi-phase description of the ISM is taken into account a
decoupling of the different thermal phases can be realized (some times
arbitrarily) and the overcooling problem can be avoided \cite{scan06,
  mura10}.

Another possibility to overcome the overcooling problem is the use of
radiative feedback schemes \cite{hqm12}.  Radiative feedback is very
relevant because it has been recently suggested that photo-heating and
radiation pressure are the most important sources of feedback in DGs
\cite{hopk13}.  Eventually, also cosmic rays have been suggested as an
additional source of feedback \cite{jube08, db12, booth13}.  Also a
correct inclusion of stellar dynamics can be a way to avoid the
overcooling problem (see Sect. 3).  A much broader discussion would
deserve the description of the feedback from the central AGN.  This
kind of feedback has gained popularity in the last decade.  It appears
in fact to be a useful recipe to use in semi-analytical models of
structure formation \cite{crot06}.  However, it is not clear how
significant the AGN feedback can be for the evolution of low-mass
galaxies.  Scaling relations \cite{fm00, mcco11} indicate that DGs
possess very small central massive black holes.  It is very likely
that all these forms of feedback occur in real galaxies.  However,
before implementing them in simulations, one should be confident that
the underlying physics is well understood and that reasonable
parametrisations can be used.

Although feedback schemes are widely debated in the literature, less
problematic appears to be the amount of energy a SN explosion deposits
into the ISM.  A value of 10$^{51}$ erg is assumed as it represents
the typical SN explosion energy $E_{\rm SN}$ \cite{ss00, ds12}.  It is
however worth reminding that SN kinetic explosion energies
(theoretically calculated or deduced from observations) cover a very
broad range, from a few 10$^{48}$ ergs of the faintest SNe to the
10$^{52}$ ergs or more of the hypernovae \cite{nomo09}. 

Some authors adopt a thermalization efficiency $\epsilon_{\rm SN}$, in
order to account for the radiative energy losses during the early
phases of the evolution of a SN remnant.  A commonly adopted value of
$\epsilon_{\rm SN}$ is 0.1 \cite{sht97}.  Indeed, the simulations of
Thornton et al.  \cite{thor98} suggest that only $\sim$ 10\% of the SN
explosion energy can be used up to thermalize the ISM.  However,
detailed simulations of the impact of isolated stars on the ISM
\cite{fhy03, krog06, hens07} show that the energy transfer efficiency
can be even lower than 1\%.  A different approach, where the
contribution of a whole population of stars is considered \cite{mg04}
clearly shows that $\epsilon_{\rm SN}$ must be a function of time.
During the early phases of galactic evolution, the SN remnants expand
in a very dense and cold ISM.  SN remnants evolve in isolation and
radiative losses are very large.  Only a small fraction of the SN
explosion energy goes to increase the thermal budget of the ISM.  When
the ISM becomes hotter and more porous, radiative losses are less
significant.  Various SN remnants quickly coalesce and form a
superbubble.  Within this superbubble, the sound speed is large.  If a
SN explodes inside the superbubble, the time it takes for the SN shock
velocity to become equal to the sound speed is very short.  This is
the time at which the shock loses its identity and the energy of the
SN remnant can be transfered to the ISM.  Clearly, in this situation
the SN remnant does not have time to radiate away a large fraction its
energy, which can be thus efficiently converted into thermal energy of
the ISM once the SN shock velocity becomes equal to the local sound
speed.

Simple analytical estimates of the thermalization efficiency as a
function of the ambient density and temperature are possible
\cite{lask67, spit68, bmd98, rmd01, stri04}.  Again, these formulae
show that $\epsilon_{\rm SN}$ is strongly reduced if the ambient
density is large and the temperature is low.  A more quantitative
evaluation of $\epsilon_{\rm SN}$ for a single, isolated galaxy can be
obtained as follows.  The stalling radius $R_s$ is defined as the
radius at which the expansion velocity of the SN shock equals the
local sound speed.  At this radius, the material inside the SN shock
can be causally connected with the external ISM and a transfer of
energy can occur.  $R_s$ can be evaluated as \cite{cmb88}:
\begin{equation}
R_s \simeq 4.93 R_{\rm pds} \left( \frac{E_{\rm SN}^{1/14}n_0^{1/7} Z^{3/14}}
{c_s}\right)^{3/7}.
\label{eq:rs}
\end{equation}
Here, the SN explosion energy is expressed in units of 10$^{51}$ ergs,
the ambient density $n_0$ in cm$^{-3}$, the metallicity $Z$ in units
of the solar metallicity and the sound speed $c_s$ in units of 10$^6$
cm s$^{-1}$.  $R_{\rm pds}$ is the radius of the SN shock at the
moment in which cooling becomes important.  Assuming that most of the
SN energy at this stage is in the form of kinetic energy of the shell
the energy available to thermalize the ISM is:
\begin{equation}
E_{\rm kin} = \frac{2}{3} \pi R_s^3 \rho_0 c_s^2.
\end{equation}
\noindent
The thermalization efficiency is now simply the ratio between this
residual energy and the initial explosion energy $E_{\rm SN}$.  Using
the value of $R_{\rm pds}$ given by \cite{cmb88}, one obtains:
\begin{equation}
\epsilon_{\rm SN} \simeq 0.02 E_{\rm SN}^{-5/98}n_0^{-54/49}
Z^{-15/98}c_s^{5/7}.
\end{equation}
\noindent
This calculation is surely approximate.  In particular, the ISM
porosity and the possibility that various SN remnants merge have not
been taken into account.  However, additional corrections could be
included and a more physically motivated description of the
thermalization efficiency, depending on the local thermodynamical
conditions, could be obtained.  

Eventually, the expansion of ionisation fronts could be taken into
account, as well.  Simple formulae could be devised to describe the
variation of the Str\"omgren radius surrounding a single massive star
or an association of stars \cite{spit78, lsp01}.  Within this radius
the cooling is indeed strongly suppressed because Ly continuum photons
are used up on the spot to ionise hydrogen atoms and only photons from
the Balmer series onwards can leave the H\,{\small II} region.
Combining these formulae with the ones describing the evolution of SN
shocks and winds from massive stars seems to be theoretically
possible.  This method is perhaps a further viable solution of the
overcooling problem.  Of course, once radiative hydrodynamical codes
will have enough resolution to solve individual H\,{\small II} regions
and SNeII remnants, these analytical considerations will be
superfluous.  However, this seems not to be possible in the
foreseeable future.

To finish this section, it is important to remind that the rate of
energy release from SNe and stellar winds is as important in galaxy
simulations as the way this energy is converted into ISM energy.  It
is commonly assumed that all the stars with masses larger than a
certain threshold mass $m_{\rm thr}$ explode as SNeII at the end of
their lifetimes.  This assumption, together with prescribed stellar
lifetime functions, makes the calculation of SNII rates quite
straightforward.  Two sources of uncertainty must be outlined.  One is
the stellar lifetime function, which is still quite uncertain and
model-dependent.  Romano et al.  \cite{roma05} demonstrated however
that uncertainties in the lifetimes of massive stars are not so
significant and do not crucially affect the results of galaxy
evolution models.  More critical is the choice of $m_{\rm thr}$.  A
commonly adopted value is 8 M$_\odot$ but, since there is still not
much known about the fate of stars in the mass interval [8:12]
M$_\odot$, $m_{\rm thr}$ could be as high as 12 M$_\odot$.  For a
Salpeter IMF extending until 100 M$_\odot$, $\sim$ 78 \% more SNeII go
off if $m_{\rm thr}=8$ M$_\odot$ instead of $m_{\rm thr}=12$
M$_\odot$.  Clearly, this is a non-negligible fraction.

Even more uncertain and less standardised are the feedback recipes
from stellar winds and Type Ia SNe (SNeIa).  Many authors even neglect
these energy contributions.  However, the energetic input of stellar
winds is very important to establish self-regulation in the star
formation process (K\"oppen et al. \cite{kth95}, see also Sect. 4).
Many authors take into account stellar winds, either adopting suitable
parametrisations based on observations \cite{tbh92}, or adopting the
results of models such as Starburst99 \cite{leit99}, which give the
mechanical energy from stellar winds released by a single stellar
population or due to a continuous episode of star formation.  This
approach has been followed, for instance, by \cite{ss00, recc07}.
Since the stellar wind luminosity decreases with metallicity
\cite{lc99, kp00}, neglecting stellar winds is perhaps acceptable in
simulations of very metal-poor DGs.

Type Ia SNe play a very important role in the evolution of galaxies,
as they are the major contributors of iron, a widely used metallicity
proxy \cite{matt12}.  Since the lifetime of SNeIa progenitors can be
as long as many Gyrs \cite{wh12}, they represent a source of energy
more evenly distributed in time than SNeII.  The relevance of SNeIa
for the dynamical evolution of galaxies has been shown for instance by
Recchi \& Hensler \cite{rh06}.  Many papers neglect the contribution
of SNeIa as they are interested in the early evolution of galaxies and
SNeIa are not assumed to occur on short timescales \cite{mf99}.
However, evidence is mounting \cite{mr01, sb05, mdp06, matt09} that a
significant fraction of SNeIa explode on timescales shorter than 100
Myr.  Thus, SNeIa should be considered in chemo-dynamical models even
if the time-span of the simulation is of the order of 100 Myr. 

A convenient parametrisation of the SNeIa rate is \cite{rc98, greg05}:
\begin{equation}
R_{Ia}(t) = \int_{t_{\rm min}}^t A_{Ia}(t-\tau) D (t-\tau) \psi (\tau) 
d\tau,
\label{eq:iarate}
\end{equation}
\noindent
where $t_{\rm min}$ is a suitably chosen minimum timescale for the
occurrence of SNeIa (typically 30 Myr), $A_{Ia}$ is a normalisation
constant and $D$ is the so-called delay time distribution (DTD), i.e.
the distribution of time intervals between the birth of the progenitor
system (usually a binary system made of two intermediate-mass stars)
and the SNIa explosion.  According to Eq. \ref{eq:iarate}, the DTD is
thus proportional to the SNIa rate following an instantaneous burst of
star formation.  Unfortunately, the form of the DTD is still very
uncertain, although some observations \cite{tota08, mmb12} suggest the
DTD to be inversely proportional to the elapsed time, i.e.  $D (t)
\propto t^{-1}$.  Studies of the chemical evolution of galaxies have
been performed \cite{matt06, matt09, bona13}, showing that the
adoption of different DTDs drastically changes the outcome of the
simulations.  It is not difficult to imagine that even more drastic
differences could be obtained in chemo-dynamical simulations of
galaxies.  The role of various DTDs on the evolution of galaxies is
another aspect that has been barely considered so far in
chemo-dynamical simulations and that, perhaps, deserves more
attention.

\section*{8. Environmental effects}
Galaxies are sociable entities; galaxies out there on their own are
quite rare.  Most of them are found in galaxy clusters and groups.  In
order to fully understand the evolution of galaxies, the study of the
galactic environment is thus paramount.  The environment not only
includes neighbouring galaxies, but also the tenuous gas between
galaxies (the intergalactic medium, IGM, or intra-cluster medium, ICM
in cluster environments).  There are many reasons why the study of
galaxy interactions and mergers is very important for our
understanding of the Universe as a whole.  Perhaps one of the most
important ones is that the largely accepted cosmological model, a
$\Lambda$ dominated Cold Dark Matter based Universe, explicitly
predicts that galaxies should form hierarchically in the merger
process.  However, the theoretical study of interactions and mergers
is usually the realm of cosmological simulations and I refer the
readers to the many books and review papers devoted to the argument
\cite{bs99, stru06, cons07, smith10, bour11, fw12}.

One of the clearest evidences of the environmental effects is the
morphology-density relation \cite{dres80}, according to which the
fraction of early-type galaxies in clusters increases with the local
density of the environment.  Another key observational result is the
star formation-density relation \cite{balo04, tana04}, in the sense
that star formation seems to be strongly reduced in dense
environments.  Moreover, cluster galaxies are H\, {\small I} deficient
compared to their field counterparts. The deficiency increases towards
the cluster centre.  These and other observational facts (see also
\cite{bg06, igle12} for reviews) clearly indicate that one or more
processes in cluster and group environments remove gas from galaxies
or make them consume their gas more quickly.  

One possibility is that the dense environment promotes tidal
interactions (galaxy-galaxy or galaxy-cluster).  It has been shown
that these interactions can remove matter from galactic halos quite
efficiently \cite{sb51, merr83, vj90, bv90}.  Another possible
physical mechanism able to remove gas in dense environment is the
combined effect of multiple high-speed encounters with the interaction
of the potential of the cluster as a whole, a process that has been
named ``harassment'' \cite{moore96, mlk98}.  The first harassment
simulations specifically targeting DGs have been performed by
Mastropietro et al. \cite{mast05}.  In this paper, it is shown that
the majority of galaxies undergo significant morphological
transformation, and move through the Hubble sequence from late-type
discs to dwarf spheroidals.  Less dramatic are the effects of
harrassment in computer simulations of late-type, disk DGs in the
Virgo Cluster \cite{smithr10}.  Strong tidal encounters, that can
morphologically transform discs into spheroidals, are rare.  They
occur in $\sim$ 15\% of infalls for typical DG orbits in the potential
of the Virgo Cluster.  Harrassment might have some impact on the
globular cluster systems of DGs, too \cite{smith13b}.

By combining different processes, Boselli \& Gavazzi \cite{bg06}
concluded that the most probable mechanism able to explain the
observational differences between galaxies in clusters and in the
field is ram-pressure stripping, namely the kinetic pressure that the
ICM exerts on the moving galaxies.  If the ram-pressure is larger than
the restoring gravitational force (per unit surface) acting on a gas
parcel of a galaxy moving through the ICM, this gas parcel is stripped
off the galaxy \cite{gg72}.  There have been many simulations
exploring the effect of ram-pressure stripping, with different
settings and degrees of sophistication \cite{fs80, amb99, ss01, rh05,
  maye06, rb06, kron08, ph12, smith13}.  There are many indications
that ram-pressure stripping is a key process, able to radically modify
the evolution of DGs.  It is interesting to note that in dwarf
irregulars, the removal of the gas by means of ram-pressure stripping
can change the potential surrounding the stars enough to dynamically
effect them, causing disk thickening by a factor of $\sim$ 2, and disk
distortion.  Actually, even the dark matter can be dynamically
effected by this \cite{smith12}.  Many authors even put forward the
idea that ram-pressure stripping can convert gas-rich DGs into
gas-poor ones.  These ideas are comprehensively summarised in many
excellent reviews \cite{sb95, maye10, hens12, skil12} and I refer the
reader to these reviews for further details.

For the purposes of this review paper, it is more convenient to
briefly summarise the results of the simulations of Marcolini and
collaborators \cite{mbd03, mbd04}.  These authors performed simulation
of flattened, rotating DGs subject to ram-pressures typical of poor
galaxy groups.  Interestingly, despite the low values of the
ram-pressure, some DGs can be completely stripped after 100-200 Myr.
However, regions of very large surface density can be found at the
front side of DGs experiencing ram-pressure stripping.  This enhanced
density can easily lead to a burst of star formation.  If the DG
experiences a galactic wind (see also Sect. 9), several parameters
regulate the gas ejection process, such as the original distribution
of the ISM and the geometry of the IGM-galaxy interaction.  Contrary
to the ISM content, the amount of the metal-rich ejecta retained by
the galaxy is more sensitive to the ram-pressure action. Part of the
ejecta is first trapped in a low-density, extraplanar gas produced by
the IGM-ISM interaction, and then pushed back on to the galactic disc.
Clearly, the interplay between galactic winds and environment is quite
complex and very few studies address this issue in detail (see however
\cite{schi05}).  This is another research field in which, in my
opinion, more can be done.  In particular, results of small-scale
detailed simulations of individual galaxies could be used in
large-scale simulations of galaxy clusters and groups, where the
interaction processes between individual galaxies and the ICM cannot
be appropriately resolved.  This is for instance the approach followed
by Creasey et al. \cite{ctb13}, who simulate the feedback effect of
SNe in a single galaxy in order to improve sub-grid models of feedback
in large-scale simulations.  This approach should perhaps be further
extended.  Also simulations like the ones of Marcolini et al. (or
similar ``wind tunnel'' experiments) could be used to better constrain
the galactic wind-ICM interactions and improve galactic cluster-scale
simulations.

\section*{9. Galactic winds}
Galactic winds are streams of high speed particles often observed
blowing out of galaxies.  They are also thought to be the primary
mechanism by which energy and metals are deposited into the
intracluster and intergalactic medium (see also Sect. 8).  Local
example of galactic winds are NGC1569 \cite{west07}, NGC253
\cite{mats09}, NGC6810 \cite{stri07} and, of course, the archetypal
galactic wind in M82 \cite{cont13}.  There is clear evidence for
galactic winds in the spectra of several z $>$ 1 galaxies
\cite{lowe97}.  Probably, the fraction of galaxies experiencing
galactic winds was larger at high redshifts \cite{pett01, pett02,
  lund12}.  A review of many observational (and theoretical) aspects
of galactic winds is given in Veilleux et al. \cite{vcb05}.

The mechanical feedback from SNe and stellar winds is the most
probable driver of galactic winds in DGs, although other mechanisms,
such as radiation pressure and cosmic rays, are possible and have been
put forward \cite{bmw91, mmt11, sns11, uhlig12, db12, hopk13}.  There
is a large (and ever growing) number of hydrodynamical simulations of
galactic winds in the literature \cite{hqm12, uhlig12, barai13, rv13}.
Many of them, especially in the past, targeted specifically DG-sized
objects \cite{dg90, dh94, stt98, db99, mf99, mfm02}.  A quite common
outcome of these simulations is that the energy deposited by SNe and
stellar winds creates large bubbles of hot, highly pressurised gas.
This gas pushes the surrounding ISM and, under favourable conditions,
a large-scale outflow can emerge.  If the outflow velocity is large
enough, the gas entrained in it leaves the parent galaxy.  A galactic
wind has been created.  If instead the wind velocity is not high
enough, the gravitational pull eventually prevails and a galactic
fountain is formed instead.  Galactic fountains are more likely in
large spiral galaxies like our own Milky Way and have been also
extensively studied in the past \cite{tt79, breg80, meli08, srm08,
  bb13}.  Given the more reduced gravitational pull, galactic winds
are more likely than galactic fountains in DGs.  The threshold
velocity for the formation of a galactic wind is typically set equal
to the escape velocity.  However, one should be aware that the motion
of gas parcels in galactic winds is not ballistic and the escape
velocity can give only an order-of-magnitude estimate of the velocity
required to escape the galactic potential well.

Many authors \cite{lars74, vader86, ds86} have speculated that, since
the binding energy of typical DGs is equal to the explosion energy of
just a few SNe, galactic winds can occur very early in DGs and can
even lead to a quick transition from gas-rich to gas-poor DGs.
However, there are three clear failings of this scenario: $(i)$ it
fails to explain the observed morphology-density correlation (see
Sect. 8), $(ii)$ it fails to explain the fact that all observed
gas-poor DGs of the Local Group possess a large fraction of
intermediate-mass stars (see \cite{mateo98, tht09} for reviews on
stellar populations of Local Group DGs), $(iii)$ if the galactic wind
occurs very early, Type Ia SNe do not have time to enrich the ISM (see
Sect. 8).  Since Type Ia SNe are the major sources of iron, one would
expect very high [$\alpha$/Fe] ratios in the stars of DGs.  Exactly
the contrary is observed: stellar populations in DGs are characterised
by very low [$\alpha$/Fe] ratios \cite{thom05, tht09}.  Indeed, many
simulations of the development of galactic winds in DGs cited above
agree on the fact that the fraction of ISM ejected out of a galaxy as
a consequence of a galactic wind must be low.  An excellent and still
very relevant review about the effect of galactic winds in DGs is
given by Skillman \cite{skil97}.

However, hydrodynamical simulations of DGs showed that the
galactic winds are often able to expel a large fraction of metals,
freshly produced during the star formation activity.  This is mostly
due to the fact that, if the initial DG gas distribution is
flattened (as observed in gas-rich DGs), the galactic wind
will preferentially expand along a direction perpendicular to the disk
(the direction of the steepest pressure gradient, see also below).
Most of the disk gas is not affected by the galactic wind.  On the
other hand, the freshly produced metals can be easily channelled along
the funnel created by the galactic wind.  Several papers in the
literature have attempted to quantitatively address this point and
study the effect of galactic winds on the circulation and
redistribution of metals in DGs.  The main results of the
often-cited work MacLow \& Ferrara \cite{mf99} are that, even in the
presence of a strong galactic wind driven by SNeII, the ejection
efficiency of unprocessed gas is almost always close to zero.  It is
different from zero only for the smallest considered galaxies (due to
their very shallow potential well).  On the other hand, the ejection
efficiency of freshly produced heavy elements is almost always close
to one.  Silich \& Tenorio-Tagle \cite{stt98} found instead that
galactic winds do not develop in most of the models, mainly due to the
presence of a hot gaseous halo surrounding the galaxy.  The effect of
off-centred SN explosions and SN explosions distributed over most of
the disk was also studied in the literature \cite{frag04}.  Metal
ejection efficiencies are reduced in this case, due to more efficient
cooling.  Wind efficiencies are found to be low even if SN is injected
directly into supersonic turbulence \cite{sb10}.  

The ejection efficiencies of individual chemical elements was
investigated, too \cite{rmd01}.  As a consequence of very short
starbursts, galactic-scale outflows carry out of the galaxy mostly the
chemical elements produced by dying stars during the most recent
episodes of SF, with large escape fraction of metals with delayed
production, like Fe and N (see also \cite{tb13}).  In fact, a
significant fraction of $\alpha$-elements, quickly produced by SNeII,
mix locally before the development of a galactic wind (see also
\cite{rh03}).  Metals produced by SNeIa and intermediate-mass stars
can be instead easily channelled along the already-formed galactic
wind and do not suffer much mixing with the walls of the wind.  The
situation is much less clear-cut in the presence of multiple bursts of
star formation \cite{recc02} or of complex SFHs \cite{recc04}.  One
should be aware of the fact that turbulence can play a decisive role
in the process of mixing metals, a mechanism usually called turbulent
mixing \cite{dimo05}.  However, it is a considerable experimental,
theoretical, modelling, and computational challenge to capture and
represent turbulent mixing and not much has been done in this
direction for astrophysical flows (but see \cite{bk05, pan08, ps10}).

An estimate of the probability of the development of a galactic wind
can be obtained as follows (see \cite{mm88, ft00, rmd01}).  Take for
simplicity a source of energy producing a constant luminosity $L$.
Assume also that the density and the metallicity of the ISM is uniform
and that its vertical density distribution has a scale height $H$.
The energy input creates a superbubble which is assumed to be
spherical and characterised by a radius $R$.  By means of standard,
textbook formulas for the evolution of a superbubble without radiative
losses (i.e. $R\sim t^{-3/5}$), the time for the radius $R$ of the
superbubble to reach $H$ is readily calculated:
\begin{equation}
t_{\rm D} \sim H^{5/3} \left( \frac{\rho}{L}\right)^{1/3}.
\label{eq:td}
\end{equation}
\noindent
However, radiative losses, in general, can not be neglected.  The
radiative losses of the hot cavity can be more relevant for the
dynamics of the superbubble than the radiative losses of the shocked
material.  The cooling timescale of the superbubble can be estimated
as:
\begin{equation}
t_{\rm c} \sim 16 (\beta Z)^{-35/22}L^{3/11} n^{-8/11}\;{\rm Myr},
\label{eq:tc}
\end{equation}
\noindent
where $L$ in this formula is in units of 10$^{38}$ erg s$^{-1}$ and
$n$ in cm$^{-3}$.  Here, $\beta$ is a numerical factor (of the order
of unity) that takes into account the fact that the cooling gas might
be out of ionisation equilibrium.  Clearly, if $t_{\rm c}$ is much 
shorter than $t_{\rm D}$, the superbubble loses much of its pressure 
before the supershell can reach $H$ and a large-scale outflow can not 
occur.  By combining Eqs. \ref{eq:td} and \ref{eq:tc} one obtains an 
approximate criterion for the occurrence of a galactic wind, namely:
\begin{equation}
L \gg 0.03 n^{7/4} (\beta Z)^{21/8} H^{11/4}.
\end{equation}
\noindent
Although this derivation is quite approximate, the large dependence of
the threshold luminosity on $H$ is a solid result.  The vertical
distribution of gas strongly affects the development of a galactic
wind (more than other factors).  A galaxy characterised by a very thin
disk experiences outflows much more easily than a roundish galaxy.
This result matches the physical intuition that in flat galaxies a
large-scale outflows easily develops along the direction of steepest
pressure gradient (i.e. perpendicularly to the disk), whereas in
spherical galaxies the pressure gradient is isotropic and either the
outflows occurs along all directions, or the superbubble remains
confined inside the galaxy.  Indeed, simulations of spherical (or
almost spherical) DGs have shown that it is not easy to create
galactic winds, even if the energy input is significant \cite{marc06}
or the galaxy does not have a dark matter halo \cite{htg04, recc07}.
Although the importance of the disk thickness for the development of
outflows was soon recognised, this aspect has not been fully explored
in the past in numerical investigation (but see \cite{ss00, stt01,
  mich07, vvs08, rha09, schr11}).

\begin{figure*}[t]
\resizebox{17cm}{!}{\includegraphics{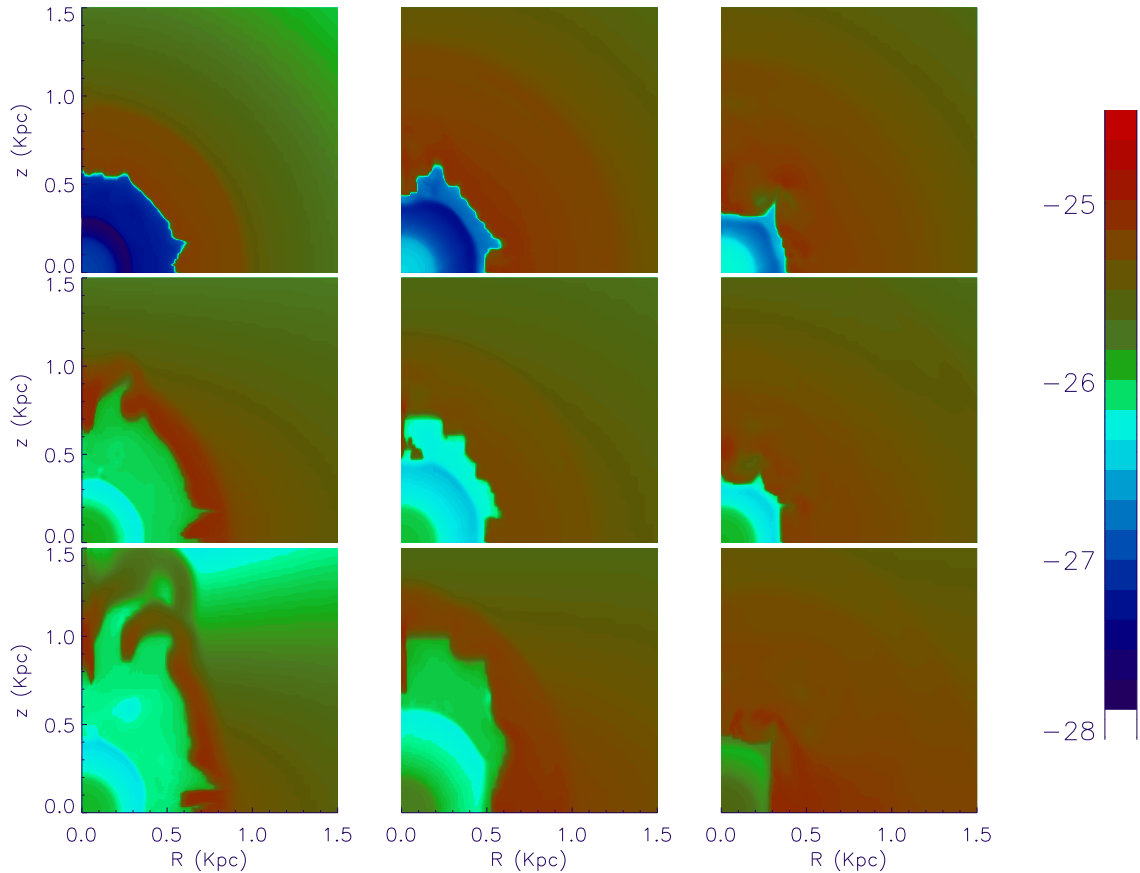}}
\caption{Gas density distribution for nine galaxy models differing on
  the degree of flattening and the initial baryonic mass, after 100
  Myr of galactic evolution.  The first column represents models with
  10$^7$ M$_\odot$ of initial baryonic mass, the middle column shows
  the gas distribution for models with mass 10$^8$ M$_\odot$, and the
  right-hand column displays the models with 10$^9$ M$_\odot$. The top
  rows of models are characterised by a roundish initial distribution.
  The middle rows show models with an intermediate degree of
  flattening, and the bottom rows are characterised by a flat initial
  distribution.  The left-hand strip shows the (logarithmic) density
  scale (in g cm$^{-3}$).}
\label{fig:covernew}
\end{figure*}

In Recchi \& Hensler \cite{rh13} we specifically addressed the role of
gas distribution on the development of galactic winds and on the fate
of freshly produced metals.  We found that the gas distribution can
change the fraction of lost metals through galactic winds by up to one
order of magnitude.  In particular, disk-like galaxies tend to lose
metals more easily than roundish ones.  In fact, the latter often do
not develop galactic winds at all and, hence, they retain all the
freshly produced metals.  Consequently, the final metallicities
attained by models with the same mass but with different gas
distributions can also vary by up to one dex.  

Confirming previous studies, we also show that the fate of gas and
freshly produced metals strongly depends on the mass of the galaxy.
Smaller galaxies (with shallower potential wells) more easily develop
large-scale outflows, so that the fraction of lost metals tends to be
higher.  An example of the results of these investigations is given in
Fig.  \ref{fig:covernew}.  The gas density distribution for nine
galaxy models differing on the degree of flattening and the initial
baryonic mass, after 100 Myr of galactic evolution is shown in this
figure (see figure caption for more details).  The effect of geometry
on the development of galactic winds is clear from this figure: the
density distribution in the models in the bottom row (flat models) is
clearly elongated.  In one case a galactic wind is already blowing.
The models in the upper row are instead still roundish.  Clearly, as
described before, if a large-scale outflow is formed, freshly produced
metals can be easily lost from the galaxy.  Any time a galactic wind
is formed, the ejection efficiency of metals is larger (some times
much larger) than the ejection efficiency of the ISM, confirming that
galactic winds must be metal-enhanced.  The fact that galactic winds
are metal-richer than the global ISM has been observationally verified
\cite{mkh02, owb05}.

\begin{figure}[t]
\resizebox{8.5cm}{!}{\includegraphics[angle=270]{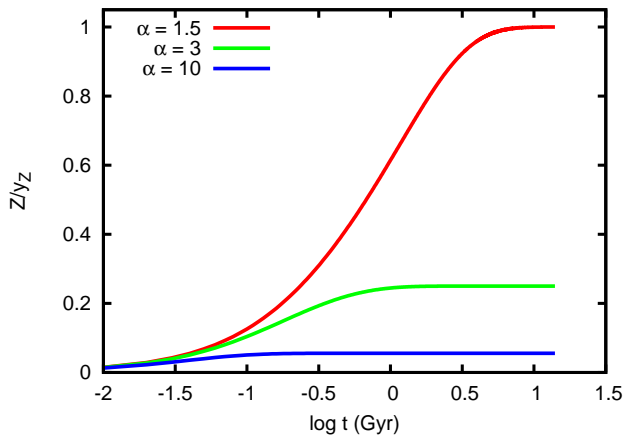}}
\caption{The metallicity of a galaxy, as a function of time, for
  models with metal-enriched galactic winds.  This plot shows the
  solution of Eq. \ref{eq:zan} for three different values for the
  enrichment parameter $\alpha$.}
\label{fig:alpha}
\end{figure}

The fact that the galactic winds are metal-enriched is a commonly
accepted result.  It has been proposed as one of the main mechanisms
leading to the so-called mass-metallicity relation, according to which
the metallicity of a galaxy grows with its mass.  Since galactic winds
are metal-enhanced and since DGs experience more easily galactic
winds, clearly one has to expect that DGs are metal-poorer than larger
galaxies \cite{trem04, spit10}.  Although the effect of metal-enriched
galactic winds on the chemical evolution of galaxies might be already
clear from the previous paragraphs, a more quantitative analysis can
be performed, based on simple analytical considerations.  Assuming
linear flows, i.e. assuming that infall rates and outflow rates in and
out of galaxies are proportional to the SFR $\psi$, a set of
differential equations can be found for the time evolution of the
total baryonic mass $M_t$, total gas mass $M_g$ and total mass in
metals $M_Z$ within a galaxy (see \cite{matt01,recc08}):
\begin{equation}
  \begin{cases}{d M_{t} \over d t} = (\Lambda - \lambda) (1 - R) \psi (t) \cr
    {d M_g \over d t} = (\Lambda - \lambda - 1) (1 - R) \psi (t) \cr
    {d M_Z \over d t} = (1 - R) \psi (t) [\Lambda Z_A + y_Z - 
    (\lambda\alpha + 1) Z]\end{cases}
\label{eq:system}
\end{equation}
\noindent

Here, $\Lambda$ and $\lambda$ are proportionality constants relating
the SFR to the infall and outflow rate, respectively.  $Z_A$ is the
metallicity of the infalling material and $R$ is the fraction of the
considered stellar populations locked into long-living stars and
remnants.  $y_Z$ is the stellar yield, in this case defined as the
ratio between the mass of a specific chemical element ejected by a
stellar generation and the mass locked up in remnants (\cite{tins80},
see also Sect. 6).  Finally, $\alpha$ is the parameter that takes into
account metal-enriched galactic winds, i.e. is the increase of
metallicity of the wind compared to the ISM.  Besides this last
factor, the equations are standard, textbook equations for the
simple-model evolution of a galaxy \cite{tins80, pagel97, matt12} and
analytical solutions can be found.  An analytical solution can be
found even including this further factor $\alpha$ (see Recchi et al.
\cite{recc08}, eq. 12).  If one assumes that the SFR $\psi$ is
proportional to the total gas mass $M_g$ through a proportionality
constant $S$ (see Sect. 4), the final result is:

\small
\begin{align}
&{Z (t) \over {y_Z+\Lambda Z_A}} = 
{{1 - \bigl[\bigl(\Lambda - \lambda + 1\bigr) - 
\bigl( \Lambda - \lambda\bigr) e^{h(t)}
\bigr]^{{\Lambda + (\alpha - 1) \lambda} \over {\Lambda - \lambda - 1}}}
\over 
{\Lambda + (\alpha - 1) \lambda}}\notag\\
& h (t) = (\lambda + 1 - \Lambda) (1 - R) S t.
\label{eq:zan}
\end{align}
\noindent
\normalsize

This solution has been plotted in Fig. \ref{fig:alpha} for
$\Lambda=0$, $\lambda=2$, $S=1$ Gyr$^{-1}$ and $R=0.26$ (from
\cite{ww95}).  The strong effect of $\alpha$ (a factor of $\sim$ 20)
on the final metallicity of the galaxy is evident from this figure.
Clearly, this kind of modelling can only give an approximate idea
about the chemical evolution of galaxies and that full chemo-dynamical
simulations are required for a deeper insight and understanding of the
metal enrichment process.  However, this kind of analytical
calculations are nowadays quite popular, as they enlighten in a simple
way complex correlations among galaxies \cite{spit10, dfd13, lilly13}.

\section*{10. Conclusions and outlook}
In this review I presented a summary of the state-of-the-art for what
concerns the chemo-dynamical modelling of galaxies in general and of
dwarf galaxies in particular.  I have devoted one Section for
each of the main ingredients of a realistic simulation of a galaxy,
namely: $(i)$ initial conditions (Sect. 2); $(ii)$ the equations to
solve (Sect. 3); $(iii)$ the star formation process (Sect. 4); $(iv)$
the initial mass function (Sect. 5); $(v)$ the chemical feedback
(Sect.  6); $(vi)$ the mechanical feedback (Sect. 7); $(vii)$ the
environmental effects (Sect. 8). In each section, commonly adopted
methodologies and recipes have been introduced and some key results of
past or ongoing studies have been summarised.  Moreover, some key
results concerning the development of galactic winds and the fate of
heavy elements, freshly synthesised after an episode of star
formation, have been summarised in Sect. 9.

Throughout this review, I outlined topics, physical processes and
ingredients that in my opinion are not properly or adequately treated
in modern simulations of galaxy evolution. I summarise below the
topics that in my opinion deserve more attention:
\begin{itemize}
\item {\bf Inclusion of self-gravity} in building initial equilibrium
  configurations.  This is clearly an important step towards building
  more realistic initial configurations and, as described in Sect. 2,
  the difference between models with and without self-gravity can be
  extremely large.  Of course, taking self-gravity into account in
  building initial equilibrium configurations is computationally
  demanding.  However, it is clearly a necessary step in simulations in
  which the star formation process is treated in detail, as the gas
  self-gravity is the main driver of the star formation process.  Of
  course, galactic simulations in a cosmological context do not need
  any special recipe to build initial configurations.
\item {\bf Inclusion of turbulence} in galactic simulations.  There is
  no doubt that the gas in galaxies is turbulent, therefore it is
  necessary to devote more efforts to a proper modelling of
  (compressive) turbulence in galaxies.  As mentioned in Sect. 7,
  turbulence is also a key ingredient to study the process of
  circulation and mixing of heavy elements in galaxies; it thus helps
  to interpret more properly observational data, such as the ones
  obtained by means of integral field spectroscopy.  As reported in
  Sect. 3, some galactic simulations with a proper treatment of
  turbulence have been already performed.  However, in these
  simulations chemistry is usually treated in a very crude and
  approximate way.  The inclusion in these simulations of methods and
  recipes about the production and circulation of heavy metals adopted
  in other chemodynamical simulations, appears to be feasible.
  Moreover, some of the assumptions and equations used to simulate
  turbulence in the ISM are based on experimental results on
  incompressible turbulence.  A more focused study of physical
  processes and modelling of compressible turbulence in the ISM is
  arguable and I am sure that in the next years we will experience
  great progresses in this field.
\item {\bf A multi-phase, multi-fluid treatment of the ISM} in galaxy
  simulations.  Realistic simulations of galaxies should take into
  account the multi-phase nature of the ISM in galaxies and the
  complex network of reactions between stars and various gas phases.
  This has been done in some simulations, particularly thanks to the
  work of Hensler and collaborators (see e.g. \cite{hb90, tbh92,
    sht97, sc02, hth06}).  These works unveiled the complexity of true
  multi-phase simulations of galaxies.  Yet, these complex simulations
  are necessary in order to reproduce more faithfully the ISM.  Tanks
  to enormous progresses in the field of multi-phase simulations in
  other branches of physics (see e.g. the monographs \cite{staed06,
    kolev07, bren09}) I hope that we can witness a boost of true
  multi-phase, multi-fluid galactic simulations in the next years.
\item {\bf Inclusion of dust.}  As already mentioned in Sect. 6, many
  works about the chemical evolution of galaxies \cite{dwek98, zgt08,
    cpm08, pipi11, zh13} include dust and show how important this
  component is to interpret data about the chemical composition of
  galaxies.  It is very likely that the inclusion of dust can
  drastically change also the results of chemo-dynamical evolution of
  galaxies and can dramatically improve our knowledge about the
  physics of the dust-gas interaction and about the circulation of
  metals in galaxies.  In spite of useful attempts, current
  state-of-the-art numerical simulations of galaxies do not take dust
  into account (but see \cite{bekki13a, bekki13b}).  A proper
  inclusion of dust is difficult and can also lead to numerical
  problems.  However, in other branches of astrophysics some of these
  numerical issues have been solved and sophisticated simulations of
  gas-dust mixtures have been performed \cite{pm06, sss10, vanm11,
    sss12}.  It would be extremely beneficial for the astronomers
  working on simulations of galaxies to learn from these works and
  improve the treatment of dust physics and dust-gas interactions in
  galactic simulations.  It is also worth noticing that the publicly
  available Pencil Code \cite{bd02, bran03, hbd04} already includes
  relevant dust physics.  A wider use of this code for simulating ISM
  in galaxies is certainly arguable.
\item {\bf A more self-consistent treatment of the IMF.}  Recent,
  detailed simulation of the ISM with a proper treatment of the star
  formation process \cite{bate09, kkm12, bate12} are able to recover
  the main shape and features of the IMF.  In these simulations, thus,
  the IMF is not assumed a priori but is self-consistently reproduced.
  Galaxy-wide simulations do not have an adequate spatial resolution,
  therefore some simplifying assumptions about the IMF need to be
  made.  Yet, it appears to me that a lot is known about physical
  properties and mass distribution of stellar clusters in galaxies,
  and these can be used to constrain the formation mechanisms of star
  clusters in galactic simulations.  Within each clusters, the
  observationally-based maximum-mass vs. cluster mass ($m_{\rm
    max}-M_{\rm cl}$, \cite{wk06, wkp13, krou13}) relation can be used
  to link the upper stellar mass within each cluster to the cluster
  mass.  This appears to be a simple and physically motivated
  exercise, that can significantly change the outcome of a galactic
  simulation.  Finally, the full IGIMF theory as developed by Kroupa
  and collaborators (see Sect. 5 and \cite{krou13} for a review) can
  be implemented in numerical simulations.  As shown in Sect. 5 with
  two simple examples, the results can drastically change compared to
  simulations adopting an universal IMF.  In spite of some attempts
  \cite{bekki13b, bekki13c, ploe13}, almost nothing has been done in
  this field.
\item {\bf Feedback recipes.}  This is a very vibrant and active field
  of research, with new methods and implementations appearing weekly
  in the preprint archives.  However, it seems to me that some
  ingredients and topics are receiving less attention than they
  deserve.  In particular, before concentrating on methods and
  algorithms to inject energy into the ISM (the kinetic, thermal and
  radiative feedback schemes described in Sect. 7) I think one should
  be sure that all relevant sources of energy are included and
  properly treated.  In particular $(i)$ Type II SNe are always
  included but it is usually not appreciated how much the total energy
  coming from SNeII can change if the threshold mass $m_{\rm thr}$
  above which SNeII can explode is changed.  As shown in Sect. 7, a
  change in $m_{\rm thr}$ can lead to a change in total SNII energy by
  a factor of almost 2.  It is also not always appreciated how
  uncertain is the fraction of the SNII explosion energy that can
  effectively thermalize the ISM.  Some analytical estimates of this
  fraction are available in the literature (see also Sect. 7) and I
  think it could be very useful to use more often and more
  consistently these kinds of analytical estimates.  $(ii)$ Type Ia
  SNe are often neglected and, if they are considered, no systematic
  study of the dependence of the results of the simulations on the
  Type Ia SN rates is available in the literature.  This appears to be
  a simple and yet quite useful exercise.  $(iii)$ Stellar winds from
  massive and intermediate-mass stars can also contribute very
  significantly to the energy budget of a galaxy, in particular if the
  metallicity is not extremely low.  This ingredient, too, is often
  neglected or not properly considered in galactic simulations.  The
  availability of softwares like Starburst99 \cite{leit99} makes the
  inclusion of stellar winds in numerical simulations quite simple.
\item {\bf Synergy between galactic scale and cluster scale or
    cosmological simulations.}  As mentioned in Sect. 8, results of
  detailed simulations of individual galaxies could be used in
  simulations of galaxy clusters, groups or even in cosmological
  simulations, in order to improve the sub-grid recipes of these
  large-scale simulations.  In particular, details of the formation of
  galactic winds and their impact on the external intergalactic or
  intracluster medium (see Sect. 9) can be extremely beneficial in
  large-scale simulations where these effects are usually treated very
  crudely.
\end{itemize}

\section*{Acknowledgements} The Guest Editors of this special issue of
Advances in Astronomy are warmly thanked for having allowed me to
write this review paper.  Many thanks to Annibale D'Ercole and Gerhard
Hensler for a careful reading of the manuscript and for very useful
suggestions and corrections.  Many thanks to Sylvia Pl\"ockinger for
having produced Fig. \ref{fig:sylvia}.  Many thanks also to Francesco
Calura, Pavel Kroupa, Nigel Mitchell, Sylvia Pl\"ockinger, Donatella
Romano, Rory Smith, Eduard Vorobyov and Svitlana Zhukovska for having
read sections of this review and for having provided very useful
comments.  Many thanks to an anonymous referee, whose comments
improved the quality of the paper.  My wife, Sonja Recchi is finally
warmly thanked for a careful English proof-reading.

\end{document}